\documentclass[floatfix]{emulateapj}
\usepackage{rotating}
\usepackage{graphicx}
\begin{document}

\def\msun{{M$_\odot$}}

\title{Adaptive Mesh Refinement Simulations of Galaxy Formation:\\ Exploring Numerical and Physical Parameters}

\author{Cameron B. Hummels \& Greg L. Bryan}
\affil{Department of Astronomy, Columbia University,
Columbia Astrophysics Laboratory,\\
550 West 120th Street, New York, NY 10027}
\email{mailto: chummels@astro.columbia.edu}

\begin{abstract}
We carry out adaptive mesh refinement (AMR) cosmological simulations of Milky-Way mass halos in order to investigate the formation of disk-like galaxies in a $\Lambda$-dominated Cold Dark Matter model.  We evolve a suite of five halos to $z=0$ and find gaseous-disk formation in all; however, in agreement with previous SPH simulations (that did not include a subgrid feedback model), the rotation curves of all halos are centrally peaked due to a massive spheroidal component.  Our standard model includes radiative cooling and star formation, but no feedback.  We further investigate this angular momentum problem by systematically modifying various simulation parameters including: (i) spatial resolution, ranging from 1700 to 212 pc; (ii) an additional pressure component to ensure that the Jeans length is always resolved; (iii) low star formation efficiency, going down to 0.1\%; (iv) fixed physical resolution as opposed to comoving resolution; (v) a supernova feedback model which injects thermal energy to the local cell; and (vi) a subgrid feedback model which suppresses cooling in the immediate vicinity of a star formation event.  Of all of these, we find that only the last (cooling suppression) has any impact on the massive spheroidal component.  In particular, a simulation with cooling suppression and feedback results in a rotation curve that, while still peaked, is considerably reduced from our standard runs.
 
\end{abstract}

\keywords{galaxies: formation -- galaxies: evolution -- methods: numerical -- hydrodynamics}

\section{Introduction}
\label{intro}

In cosmologies dominated by Cold Dark Matter (CDM), galaxy rotation is produced by gravitational tidal torques arising from the hierarchical collapse of structure.   Analytic models and N-body simulations have shown that this can produce enough angular momentum to explain the observed sizes of disk galaxies \citep{Fall:1980p1033, Mo:1998p1034}.  However, computational models including gas dynamics have struggled to reproduce realistic disk galaxies.  Such models initially produced undersized disks with low angular momentum \citep{Navarro:1991p1002, Navarro:1994p1004}.  Later work did generate disks with approximately the correct extent \citep{Steinmetz:2002p1035, Abadi:2003p641}, but these had oversized stellar spheroidal components and therefore unnaturally large core circular velocities \citep{Steinmetz:1999p635} far in excess of those found observationally.  This \emph{angular momentum problem} remains one of the major shortcomings of the CDM paradigm.

The origin of the angular momentum problem is not entirely understood, but it probably stems from the fact that such simulations do not achieve sufficient spatial and mass resolution to correctly model the appropriate physical processes.  For example, insufficient spatial resolution leads to the spurious mixture of hot and cold components of the ISM, producing an artificial warm component which is very efficient at radiating away energy \citep{Katz:1992p1037, Katz:1996p1031, Steinmetz:1995p1005}.  Thus, underresolved gas in such simulations can cool very quickly, and in cooling it loses its pressure support and collapses into dense knots of material.  These knots interact with the galactic dark matter component through dynamical friction processes, and much of the angular momentum of the gas knots is transferred to the dark matter halo of a galaxy \citep{Donghia:2006p1032}.  Consequently, the knots tumble into the center of the galaxy to produce a dense cusp of material in the core of the simulated galaxy.  Typically the buildup of these massive cores with cold gas and stars occurs rapidly, and even nascent simulated galaxies exhibit evidence of these cusps as early as $z \sim 4$-5.

Energetic feedback from star formation events has the ability, in theory, to alleviate this problem either by heating and ``puffing" up the collapsing knots so they are more easily disrupted before losing their angular momentum \citep{Weil:1998p1030}, and/or by preferentially ejecting low angular momentum gas \citep[e.g.,][]{2010arXiv1010.1004B}.  However, existing simulations are unable to resolve the detailed structures of star-forming regions.  Individual star-forming and stellar feedback events occur on parsec scales.  Stellar feedback processes can be resolved in simulations confined to local regions of the ISM \citep[e.g.,][]{Joung:2006p1038} but not in cosmological simulations, which need to accurately co-evolve a galaxy's environment on scales of $ > 10$ Mpc.  

Therefore, cosmological simulations simplify and parameterize star formation and stellar feedback on scales more easily met by current computational resources (i.e. scales of $10^2 - 10^3$ pc).  A variety of techniques have been suggested in order to achieve this puffiness within the confines of low-resolution models.  In the most primitive prescription, stellar feedback is simply the return of energy from newly created star particles to their surrounding gas, usually in the form of thermal energy.  Star ``particles'' in these simulations typically represent clusters of stars of mass $10^{2} - 10^{5}$ \msun, so the thermal feedback is justified as the energy output from the most massive stars in the cluster becoming type II supernovae shortly after the creation of the star particle.  In nature, this supernovae heating produces a small component of very hot gas surrounding the stellar population, so hot that its cooling time is very long.  Unfortunately at the resolution currently achievable in cosmological simulations, this primitive thermal prescription for feedback deposits the same SN energy into a much larger reservoir of gas, which does not reach the same high temperature as it should.  This now-warm component of gas can easily radiate the excess energy away, cool further and proceed with star-formation runaway thus defeating the purpose of the feedback \citep[e.g.,][]{Steinmetz:1999p635}.  

Building from these failures, a number of research groups artificially turn off cooling in a gas parcel for a period of time ($t \sim 10^7$ yr) after a cluster of stars has formed out of it \citep[e.g.,][]{Gerritsen:1997p1039, Thacker:2000p1040, SommerLarsen:2003p1116, Stinson:2006p1023, Governato:2007p1022, Agertz:2010p461, Colin:2010p1053, Piontek:2011p1041, Guedes:2011p1080}.  This method is justified as an application of the Sedov-Taylor blast wave solution for a Type II SN \citep{Taylor:1950p1075, Sedov:1959p1074}, which blows out any cold media from the immediate environment of a star formation event.  Using this prescription, any gas in a galaxy which starts to collapse into knots will reach the star formation criteria, form a star, and then heat up without any allowed cooling, thus preventing further collapse.  Not surprisingly, these research groups have found some success with this method, yielding simulated galaxies with reduced inner rotation curves due to less massive bulge components; however, gas parcel masses and sizes in cosmological simulations of this sort are typically too large for the Sedov-Taylor solution to apply (see Section \ref{code_coolingsuppression}).  Thus despite the successes of the cooling suppression feedback model, the community continues to search for other more physically-motivated solutions.  

Another subgrid model for feedback (i.e. on scales smaller than the true resolution of the simulation) is to inject kinetic energy directly into the gas; this can alleviate the problem of thermal energy being radiated away.  For example, some studies \citep[e.g.][]{Springel:2003p1044, Scannapieco:2006p1118, Oppenheimer:2008bu} using Smoothed Particle Hydrodynamics (SPH) give some of the SN energy to individual gas particles in the form of momentum.  This method can result in significant mass outflows (by design), but at the cost of decoupling wind particles from hydrodynamic interaction for a period of time.  An alternate approach, to keep wind particles coupled to the disk gas was explored by \citet{Schaye:2008p1045}.  Both approaches help but, by themselves, do not appear to generate realistic rotation curves.

In addition, \citet{Truelove:1997p1046} showed that insufficient resolution in a simulation can lead to artificial fragmentation of the gas, perhaps resulting in a further overproduction of stars.  One way to prevent artificial fragmentation is to add additional (numerical) pressure in high-density, low-temperature regions to ensure that the Jeans length is always resolved \citep{Machacek:2001p1047, Robertson:2008p1017}.  This can be achieved by modifying the equation of state (EOS) itself, making it stiffer in order to provide an additional source of pressure to gas in denser regions \citep{Schaye:2008p1045, Ceverino:2009p1014, Agertz:2010p461}.  A polytropic EOS ($P \propto \rho^{\Gamma}$) with $\Gamma = 4/3$ will keep the ratio of Jeans length to resolution length constant (assuming Lagrangian resolution such that the resolution length decreases as $\rho^{-1/3}$ -- for fixed resolution $\Gamma = 2$ is required), but even stiffer relations have been used.  For example, \citet{Agertz:2010p461} ran simulations with such an equation of state, where in low-density regions it behaved as an ideal gas, but in high-density (star forming) regions it followed a polytropic equation of state with $\Gamma = 2$.

In this paper, we undertake an investigation of galaxy formation using an Adaptive Mesh Refinement (AMR) hydrodynamics code.  The majority of work in this field has used SPH codes, and so this allows us to investigate the problem from a new angle.  Although there has been some work with AMR codes \citep{Joung:2009p1010, Ceverino:2009p1014, Agertz:2010p461, Colin:2010p1053}, there has not been a clear demonstration that an equivalent AMR calculation (i.e. one without a subgrid feedback model) actually does reproduce the classic SPH result.

We begin by simulating a set of five halos without any feedback or subgrid model (except a minimum pressure support to prevent artificial fragmentation).  We find, in agreement with SPH codes that a large, concentrated bulge is produced, resulting in a rotation curve that rises to $\sim 500$ km/s at $r \sim 1$ kpc.  We then vary a number of numerical and physical parameters in order to understand how sensitive the result is to our a choice of parameters.

The paper is organized as follows.  Section \ref{method} describes the details of our hydrodynamics code, our initial conditions and the relevant parameters for this study.  In Section \ref{results}, we present the results of our simulations including the five canonical runs, our resolution study, and our modified runs.  Section \ref{discussion} is a discussion of our results and their implications.  Finally, Section \ref{summary} summarizes our conclusions and makes predictions for future solutions to the angular momentum problem.

\section{Methodology}
\label{method}

\subsection{Code}
\label{code}
Our simulations were performed using \emph{Enzo}\footnote{enzo-project.org}, an Eulerian, three-dimensional, grid-based hydrodynamics code that employs adaptive mesh refinement in order to achieve targeted regions of high resolution in a cosmological volume \citep{Bryan:1997p869, OShea:2004p446}.  Gas is discretized on the grid, but dark matter and stars are treated as particles.  The \emph{ZEUS} hydrodynamics code \citep{Stone:1992p1117} is used to evolve the gas on the grid.  \emph{Enzo} includes gas, self-gravity, a non-equilibrium model for H and He ionization and cooling, a metagalactic ultraviolet background \citep{Haardt:1996p1000}, and equilibrium cooling due to metals (although for the runs described in this paper, we do not include metal cooling).  

\subsubsection{Star Formation}
\label{code_starformation}

Star formation is modeled using a simple criteria based on \citet{Cen:1992p1071}.  A grid cell will produce a star if: (i) the overdensity in that cell exceeds a given value ($\delta_{\rm SF}$), (ii) the mass of gas in the cell exceeds the local Jeans mass, (iii) there is locally convergent flow (i.e. the velocity divergence is negative) and (iv) the cooling time for the gas to collapse is less than the dynamical time in that cell (or the temperature is near the minimum allowed, around $10^4$ K).  If a grid cell meets all the previous criteria then gas is converted into a ``star particle'', as calculated using
\begin{equation}
m_{*}=\epsilon_{\rm SF}\frac{\Delta t}{t_{\rm dyn}}\rho_{\rm gas}\Delta x^3
\label{eq:sfr}
\end{equation}
where $\epsilon_{\rm SF}$ is the star formation efficiency (more properly the efficiency per dynamical time), $\Delta t$ is the size of the time step, $t_{\rm dyn} = (3 \pi / 32 G \rho)^{1/2}$ is the dynamical time and $\rho_{\rm gas}$ is the gas density.  If the resulting star particle would have a mass above a minimum mass, M$_{* \rm ,min}$, it is formed immediately.  This mass is chosen so that a large number of small star particles will not slow down the simulation.  However, if the star particle would have a smaller mass, the probability that it will form is equal to the ratio of the mass of the projected star particle to M$_{*\rm ,min}$.  If the star particle is then formed, its mass is the minimum of M$_{*\rm ,min}$ and 90\% of the mass in the gas cell \citep[see also][]{Tasker:2006p1072}.   For all of the simulations in this study, we used a value of M$_{*\rm,min} = 10^5$ \msun.  Our default value for $\epsilon_{\rm SF} = 10^{-2}$, although this was varied in some runs (see Table \ref{tab:halos}).  We adopt $\delta_{\rm SF} = 1000$, corresponding approximately to a density threshold of 0.1 cm$^{-3}$ at $z=3$; this value simply ensures that star formation is limited to strongly non-linear regions.

\subsubsection{Resolution}
\label{code_resolution}

Because \emph{Enzo} is an AMR code, it can dynamically refine the spatial resolution of the grid when certain criteria are met in a particular region of the simulated volume.  We set the grid-refinement criteria to increase refinement whenever the dark matter mass in a cell in larger than four times the dark matter particle mass, with an equivalent criterion for the baryonic mass.  When refined, the cell resolution is increased by a factor of 2.  The placement of these refinement regions is recalculated regularly throughout the simulation, to assure that moving or emerging regions of interest are always well-resolved.  For the canonical runs, we cap this refinement when our cell sizes reach 9 levels of refinement or 425 comoving parsecs.  In addition, we conduct a resolution study in Section \ref{resolution} where we vary the resolution from 7 to 10 levels of refinement (212 - 1700 comoving parsecs).

\subsubsection{Minimum Pressure Support}
\label{code_pressuresupport}

To prevent artificial Jeans fragmentation, for most runs we implement the minimum pressure support described in \citet{Machacek:2001p1047} such that the ratio of the the Jeans length to cell size, $J = L_J / \Delta_x$, is at least 8.  We do this by adding an additional artificial pressure to the most highly refined grid cells such that this ratio is always maintained.  The addition of this pressure is intended to prevent gas clouds from collapsing below the resolution scale of the simulation, which could cause spurious numerical effects, such as artificial fragmentation.

\subsubsection{Star Formation Efficiency}
\label{code_starformationefficiency}

\begin{figure}
\begin{center}
\includegraphics[width=9cm]{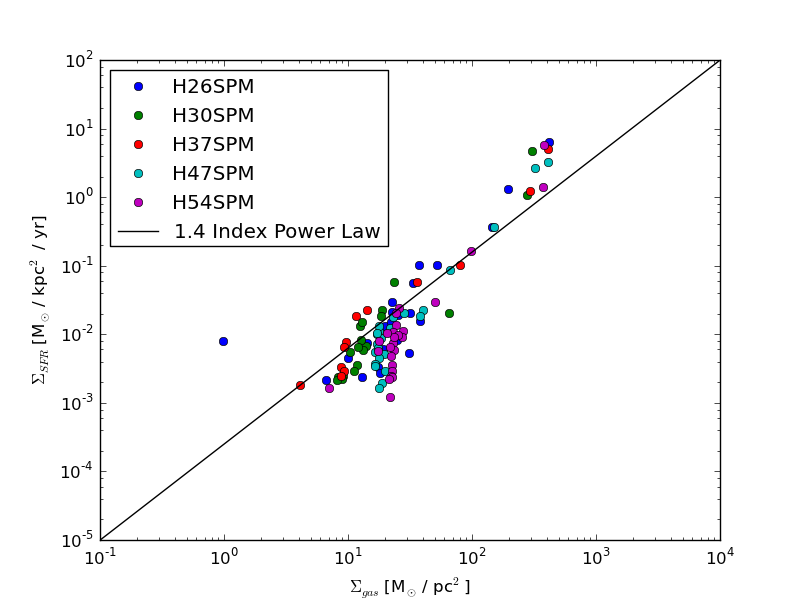}
\end{center}
\caption{We plot gas surface density versus star formation rate surface density for our five canonical runs.  We overplot the power law $\Sigma_{SFR} = 2.5\times10^{-4}\cdot \Sigma_{gas}^{1.4}$ (in these units) representing the Kennicutt-Schmidt local relation between star formation rate and gas surface density \citep{Kennicutt:1989cu}.}
\label{fig:KS}
\end{figure}

For our canonical runs, we set the star formation efficiency per dynamical time $\epsilon_{\rm SF}$ to 1\%, a parameter value that previous work found to approximately reproduce the Kennicutt-Schmidt relation for our chosen star formation law \citep{Tasker:2006p1072}.  In Figure \ref{fig:KS}, we demonstrate that our canonical simulations roughly agree with the Kennicutt-Schmidt relation by plotting the gas surface density versus the recent star formation surface density (age$_{\rm star}$ $\le$ 5 Myr) in concentric annular cylinders of 0.25 kpc width and scale height 5 kpc each aligned with their respective gas disks.  Overplot is a power-law with index 1.4, the accepted form of the Kennicutt-Schmidt law \citep{Kennicutt:1989cu}.  Our canonical value for the parameter (the star formation efficiency per free fall time) is $\epsilon_{\rm SF} = 10^{-2}$, which is in line with (but slightly lower than) typically suggested values \citep[e.g.][]{Krumholz:2007p1115}. Other galaxy formation simulations adopt different values, for example \citet{Abadi:2003p641} use a value of 3.3\%, \citet{Governato:2007p1022} use 5\%, \citet{Stinson:2006p1023} use 5\%, all for SPH simulations.  Turning to AMR simulations, \citet{Agertz:2010p461} used 1\% for their most successful runs, \citet{Teyssier:2010ia} also adopted 1\%, while \citet{Gnedin:2011ds} adopted 0.5\%, arguing that it was a better fit to the observations of \citet{Bigiel:2008bs}.  It is likely that the choice of star formation efficiency required depends both on the resolution of the simulation, as well as feedback prescription.  In this paper, we explore variations in the efficiency, decreasing this parameter to explore the impact of slowing the conversion of gas into stars.

\subsubsection{Fixed Physical Resolution}
\label{code_physicalresolution}

Most of the previous work on this topic was done using hydrodynamics codes in \emph{physical} or \emph{proper} units as opposed to \emph{comoving} units.  Our code uses comoving coordinates, which means that a fixed maximum resolution level therefore corresponds to a fixed comoving resolution at the finest refinement level (cold, dense gas is almost always refined at the maximum level in these simulations).  Thus our models resolve more details in the early universe (by a factor of $1+z$) relative to simulations employing the equivalent resolution in fixed physical units.  In an effort to explore the impact of using fixed physical rather than comoving resolution, we modified the code to be able to run with a fixed physical resolution, by adjusting the maximum refinement level with redshift.  We note that because of the discrete, factor of two changes in grid resolution, this only keeps the maximum refinement to a fixed physical resolution within a factor of two.  We do not change the refinement criteria, therefore our mass resolution remains the same (see Section \ref{initialconditions}).  Section \ref{physicalresolution} describes the results of these runs.

\begin{table*}[htb]
\begin{center}
\footnotesize{
    \begin{tabular}{|c|c|c|c|c|c|c|c|c|c|c|c|c|}
    \hline
    Run name & Halo & $M_{\rm 200}$ & $M_{\rm DM}$ & $M_{\rm star}$ & $M_{\rm gas}$ [hot, cold] & $r_{\rm 200}$ & $v_{\rm circ,max}$ & $\Delta x$ & $\epsilon_{\rm SF}$ & $\frac{\Delta x}{L_J}$ & $\epsilon_{\rm FB} $ & $ t_{\rm supp} $ \\
    & & ($10^{12}$ \msun) & ($10^{11}$ \msun) & ($10^{11}$ \msun) & ($10^{10}$ \msun) & (kpc) & (km/s) & (pc) & & & & (Myr)\\
    \hline

    H26SPM* & 26 & 1.1 & 9.4 & 1.6 & 3.7 [3.4, 0.3] & 210 & 560 & 425 & $10^{-2}$ & 8 & 0 & 0\\
    H30SPM & 30 & 1.1 & 9.1 & 1.5 & 5.1 [4.8, 0.3] & 210 & 460 & 425 & $10^{-2}$ & 8 & 0 & 0\\
    H37SPM & 37 & 0.8 & 6.6 & 1.1 & 3.4 [3.2, 0.2] & 190 & 480 & 425 & $10^{-2}$ & 8 & 0 & 0\\
    H47SPM & 47 & 0.6 & 5.1 & 0.9 & 2.1 [1.3, 0.8] & 170 & 470 & 425 & $10^{-2}$ & 8 & 0 & 0\\
    H54SPM & 54 & 0.5 & 4.0 & 0.7 & 2.0 [1.5, 0.5] & 160 & 410 & 425 & $10^{-2}$ & 8 & 0 & 0\\
    D7H26SPM & 26 & 1.2 & 9.3 & 1.8 & 4.5 [2.3, 2.2] & 210 & 440 & 1700 & $10^{-2}$ & 8 & 0 & 0\\
    D8H26SPM & 26 & 1.2 & 9.4 & 1.8 & 4.2 [2.5, 1.6] & 210 & 470 & 850 & $10^{-2}$ & 8 & 0 & 0\\
    D9H26SPM* & 26 & 1.1 & 9.4 & 1.6 & 3.7 [3.4, 0.3] & 210 & 560 & 425 & $10^{-2}$ & 8 & 0 & 0\\
    D10H26SPM & 26 & 1.3 & 11. & 1.4 & 4.8 [4.3, 0.5] & 220 & 610 & 212 & $10^{-2}$ & 8 & 0 & 0\\
    H26S & 26 & 1.1 & 9.5 & 1.6 & 3.8 [3.4, 0.4] & 210 & 540 & 425 & $10^{-2}$ & 0 & 0 & 0\\
    H26SPML & 26 & 1.2 & 9.5 & 1.7 & 4.4 [2.1, 2.3] & 210 & 640 & 425 & $10^{-3}$ & 8 & 0 & 0\\
    H26SPMR & 26 & 1.2 & 9.4 & 1.9 & 3.1 [2.5, 0.5] & 210 & 550 & 425** & $10^{-2}$ & 8 & 0 & 0\\
    H26SPMF & 26 & 1.3 & 11. & 1.4 & 6.3 [5.6, 0.8] & 220 & 580 & 425 & $10^{-2}$ & 8 & 3E-6 & 0\\
    H26SPMC & 26 & 1.2 & 9.4 & 1.7 & 8.6 [7.5, 1.0] & 210 & 400 & 425 & $10^{-2}$ & 8 & 0 & 50 \\
    H26SPMFC & 26 & 1.2 & 9.5 & 1.2 & 10. [6.5, 3.5] & 210 & 410 & 425 & $10^{-2}$ & 8 & 3E-6 & 50 \\
    H26SPMFCR & 26 & 1.2 & 9.5 & 1.1 & 11. [8.2, 2.4] & 210 & 370 & 425** & $10^{-2}$ & 8 & 3E-6 & 50 \\
    \hline
    \end{tabular}
    \footnotetext[1]{Halos H26SPM and D9H26SPM are the same simulation.} 
    \footnotetext[2]{Halos H26SPMR and H26SPMFCR have a fixed \emph{physical} resolution throughout the simulation, while all other halos have a fixed \emph{comoving} resolution.}
}
\caption{We present the bulk properties and simulation parameters for each halo and simulation.  The first five halos are the canonical simulations, each following a different halo in the same overall volume.  Additional simulations were created using the initial conditions for halo H26SPM with modifications to the encoded baryonic and star formation physics.}
\label{tab:halos}
\end{center}
\end{table*}

\subsubsection{Supernovae Feedback}
\label{code_feedback}

Most simulations we discuss in this paper do not include explicit feedback; however, we do carry out a few runs which included prompt, energetic feedback from Type II SNe.  We apply a simple prescription for stellar feedback to mimic the effects of type II supernovae.  Because individual star-formation events produce star particles of $M_{\rm star} \sim 10^{4-5}$ \msun, we spread feedback over an extended period parameterized by 
\begin{equation}
\frac{dM_{\rm form}}{dt} = M_0 \frac{t-t_0}{\tau} \exp{\frac{-(t-t_0)}{\tau}},
\label{eq:feedback}
\end{equation}
where $M_0$ and $t_0$ are the initial star particle mass and star particle creation time \citep[see also][]{Cen:1992p1071}, and $\tau$ is the maximum of either the dynamical time of the gas out of which the star particle formed, or 10 Myr (to prevent unrealistically short dynamical times).  We take a fraction of the rest mass energy $\epsilon_{\rm FB}$ of the forming stars as the available feedback energy.  This parameter can be computed assuming an initial mass function and a minimum initial stellar mass for producing a Type II SN.  

As described in \citet{Tasker:2006p1072}, we simply add the thermal energy to the local grid cell. As the surrounding gas heats up, it increases the Jeans length and theoretically quenches star formation for a time.  As was noted in Section \ref{intro}, if the resolution of the simulation is insufficient to clearly differentiate between a cold, neutral interstellar medium and a hot, ionized interstellar medium, then the pumping of this energy into the cold star-forming gas results in an unrealistic warm component (i.e. the mixing of hot and cold phases).  This warm component sits near the peak of the cooling curve of the gas, and thus effectively radiates away its energy very quickly.

For the run which does include feedback, we take a value of $\epsilon_{\rm FB} = 3 \times 10^{-6}$, which corresponds to one $10^{51}$ erg SN for every 180 \msun\ of stars formed.  The value of this parameter is uncertain, as it depends on both the initial mass function, as well as assumptions about how the energy is radiated away immediately.  Other values are used in the literature, for example \citet{Abadi:2003p641} employed a feedback prescription which corresponds to, in our definition, a value of $\epsilon_{\rm FB} = 5.6 \times 10^{-6}$.  They, like others, found that this energy was quickly radiated away, and so the value had little impact.  As we discuss in more detail below, \citet{Stinson:2006p1023} added a cooling suppression model, and argued for a value of $\epsilon_{\rm FB} = 4.3 \times 10^{-7}$, although \citet{Governato:2007p1022} adopted $\epsilon_{\rm FB} = 2.6 \times 10^{-6}$, using a very similar model.  Turning from SPH to AMR simulations, we note that \citet{Agertz:2010p461} used $\epsilon_{\rm FB} = 5.6 \times 10^{-6}$ for their standard runs (also with a cooling suppression model).  Neither \citet{Teyssier:2010ia} nor \citet{Gnedin:2011ds} appear to have used thermal feedback in their simulations.  Finally, \citet{Cen:2011bp} used $\epsilon_{\rm FB} = 10 \times 10^{-6}$ (larger than the usual efficiency because of a postulated contribution from prompt Type Ia SN); this work did manage to drive winds without a cooling suppression model, although they added the energy to the nearest 27 cells weighted inversely by density (and also had somewhat worse mass and spatial resolution than used here).  In summary, we see that our chosen value is within the range used by other researchers.

\subsubsection{Cooling Suppression Models}
\label{code_coolingsuppression}

One way to prevent this energy from being quickly lost is to turn off radiative cooling in the region immediately surrounding newly formed stars.  This was first attempted by \citet{Gerritsen:1997p1039}, but has since been explored by a range of simulations \citep[e.g.,][]{Thacker:2000p1040, SommerLarsen:2003p1116, Stinson:2006p1023, Governato:2007p1022, Agertz:2010p461, Colin:2010p1053, Piontek:2011p1041, Guedes:2011p1080}.  The idea is to use the Sedov-Taylor solution for a blast wave to model the subgrid shock physics that the code cannot resolve.  It might seem most straightforward to use the length and time-scales of the energy-conserving phase (i.e. Sedov phase) to control where and when to turn cooling off, but these turn out to be too small and too short (a few $10^4$ years) to make a significant difference, as acknowledged by previous studies \citep{Stinson:2006p1023}.  Instead, the method employs the radius and time at the end of the momentum-conserving phase under the assumption that during this phase, much of the energy is conserved in kinetic motion (and hot, diffuse gas), which the code cannot model, and would otherwise dissipate.  The larger length and timescales of the snowplow phase used are further justified as the combined forces of multiple supernovae in the star particle; however, although these supernovae may interact in a complex way, it is unlikely their effects will simply add constructively.

This prescription for cooling suppression feedback has generally been used in SPH codes, but recently some AMR codes have also adopted this technique \citep{Agertz:2010p461, Colin:2010p1053}.  The common prescription is to suppress cooling for a period of time ($30-50$ Myr) in the gas immediately around a star-formation event, regardless of where the star goes afterwards.  In our implementation, we suppress cooling of the gas in the single cell in which the star particle resides.  This is done for 50 Myr after the star particle is first created.  Since both of these length and time scales correspond closely to those over which energy from the star particle is injected in the simulation (the feedback follows Eq.~\ref{eq:feedback}, above), this acts to suppress cooling in newly heated gas.  
Given our chosen cell size (425 comoving pc in most runs), these length and timescales are similar to the region and duration of influence adopted by other researchers \citep{Stinson:2006p1023, Colin:2010p1053}.

\subsection{Initial Conditions}
\label{initialconditions}

For this work, we use the WMAP 5-year results as our cosmological parameters \citep{Komatsu:2009p1070}, in particular we use: $\Omega_{0} = 0.258$, $\Omega_{\Lambda} = 0.742$, $\Omega_{\rm baryon} = 0.044$, $\sigma_8 = 0.796$, $H_{\rm o}$ = 71.9 km s$^{-1}$.  We generate our initial conditions using \emph{inits}, a program included in the \emph{Enzo} suite.  \emph{Inits} sets up a $128^3$ particle-mesh grid, and modifies the velocity and position of the dark matter particles in each grid cell as specified by linear perturbations with the required power spectrum at $z = 99$.  The initial conditions are generated for a cubic volume of $L = 20 h^{-1}$ comoving Mpc on a side with periodic boundary conditions.  First, the simulation is populated with only dark matter particles at low-resolution ($M_{\rm DM} = 3.2 \times 10^8$ \msun) and run to $z = 0$, where candidate Milky-Way-like halos are identified.  Halos are chosen based on their final mass and accretion history, preferentially selecting halos with final masses $M_{200} \sim 10^{12}$ \msun, and those which have not undergone major mergers after $z \sim 2$ ($M_{200}$ is the mass enclosed within a radius corresponding to a mean density of 200 times the critical density).  Five such halos are identified, ranging in mass from $M_{200} = 4.8 \times 10^{11}$ to $ 1.2 \times 10^{12}$ \msun.

For each halo, the component dark matter particles are traced back to their positions in the initial conditions at $z = 99$.  This Lagrangian volume is further refined with two additional levels of refinement.  It is here that the initial conditions are regenerated, and in these nested boxes, we additionally refine the dark matter particle masses by a factor of 8 for each region.  The resulting high-resolution dark matter particle mass within the vicinity of each halo is $M_{\rm DM} = 4.9 \times 10^6$ \msun.  

A series of new simulations are performed using these new initial conditions; each one focuses on a different halo.  Baryons are included in these runs.  The high-resolution regions are further refined dynamically with adaptively-placed grids, using the refinement scheme described earlier.

\subsection{Description of Simulations}
\label{description}

We conducted five canonical simulations using identical simulation parameters and the initial conditions described  above.  These simulations are referred to as: H26SPM, H30SPM, H37SPM, H47SPM and H54SPM.  Additionally, several different runs were performed on the initial conditions for halo 26 (H26SPM), which systematically varied simulation and physical parameters to investigate the effects of each on galactic evolution.  The parameters toggled (and their respective simulations) include: (i) excluding minimum pressure support [H26S]; (ii) changing the maximum spatial resolution [D7H26SPM, D8H26SPM, D10H26SPM]; (iii) using a constant physical resolution instead of a constant comoving resolution [H26SPMR, H26SPMFCR]; (iv) including thermal feedback [H26SPMF, H26SPMFCR]; (v) lowering the star-formation efficiency [H26SPML]; and (vi) suppressing cooling in star forming regions [H26SPMC, H26SPMFC, H26SPMFCR].  The details of the various simulations and the resulting galaxies are shown in Table \ref{tab:halos}.


\section{Results}
\label{results}

In this section, we present the results of our galaxy formation simulations, first describing the five canonical runs, which all contain identical physical prescriptions but track different galactic halos.  Then, we explore variations in resolution as well as the numerical parameters we use to describe the gas and star formation.

\begin{figure*}
\begin{center}
\plotone{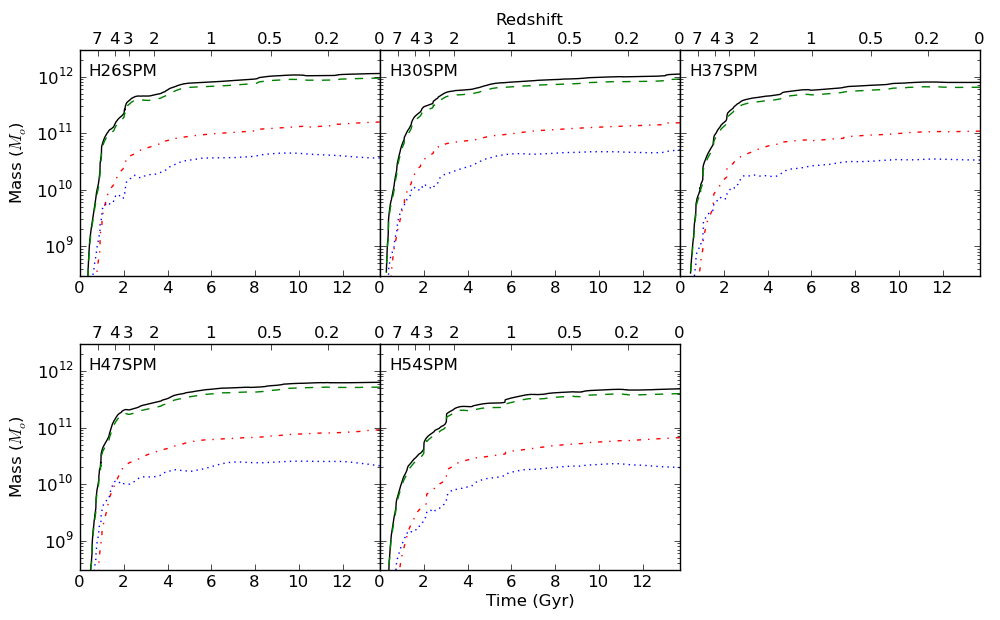}
\end{center}
\caption{We plot the mass of our five canonical halos as a function of time.  Different components of each halo are labeled differently: dark matter (green dashed line), stars (red dot-dashed line), gas (blue dotted line), and total mass (black solid line).}
\label{fig:canonical_mass}
\end{figure*}
\subsection{Mass History}
\label{masshistory}

The mass accretion history for a galaxy including the different modes of its accretion is thought to play a crucial role in determining its final dynamical state \citep[e.g.][]{Keres:2005p1111}.  In order to analyze the simulation in high time resolution, we record outputs from the simulation every 10 Myr.  For each output we run the HOP algorithm \citep{Eisenstein:1998p1073} on the dark matter particles in order to identify halos.  Given the particles in each of our five halos at $z=0$, we identify and track these halos back to early times.  Each halo is tracked backwards in time by identifying the local progenitor which shares the largest number of tightly-bound dark matter particles.  The resulting mass-accretion history for each halo is shown in Figure~\ref{fig:canonical_mass}.  The halo masses are computed inside of $r_{200}$, the radius within which the mean density is 200 times the critical density of the universe at that redshift.  At each time, we determine the center of the halo using an iterated center-of-mass technique, which starts with the center of mass within $r_{200}$, and then successively recomputes the center of mass in smaller spherical volumes, decreasing the radius by 5\% on each iteration and using the center of mass of the previous volume.  This is necessary in order to make an accurate determination of the halo center (we found that simply choosing either the densest point or the center of mass within $r_{200}$ did not produce a good estimate of the center in many cases).  All masses calculated are masses contained within $r_{200}$, and are shown in Table~\ref{tab:halos}.  

In Figure~\ref{fig:canonical_mass}, the mass histories of dark matter (green), gas (blue), stars (red) and total mass (black) are shown.  The halos are arranged in decreasing z=0 total mass.  All of the halos lack any major mergers over the last 10 Gyr, providing them with quiescent growth, in order to maximize the chance of producing disk systems.  The lowest-mass system, halo 54, undergoes mergers at $z\sim3.5$ and $z\sim2.5$  causing discrete jumps in the mass of all its components at those times.

\begin{figure*}
\begin{center}
\plotone{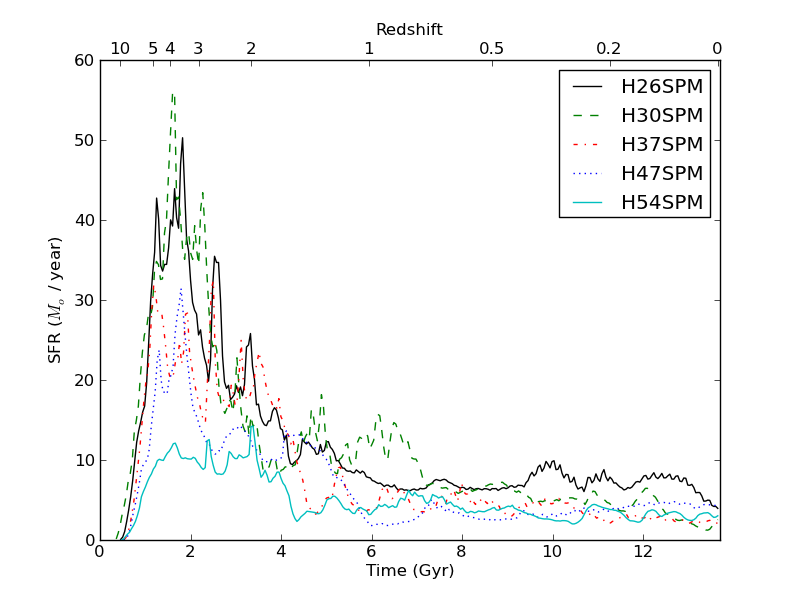}
\end{center}
\caption{We plot the star formation histories for the five canonical halos.  The bulk of the star formation occurs in the first 2-3 gigayears of evolution for each halo before settling into a low level of continuous star formation lasting to present.}
\label{fig:canonical_SFR}
\end{figure*}

\subsection{Star Formation History}
\label{starformationhistory}

Much of the angular momentum problem stems from an overproduction of stars and a buildup of the oversized stellar bulge, revealed by the galactic star formation rate history.  If one can reduce star formation early in a halo's evolution it will moderate the amount of material in the inner region of the galaxy.  While it is true that the bulk of the modeled bulge mass is stellar in nature, it remains unclear as to where these stars were created.  Were they formed in clumps in the disk before plummeting to the center of the system, or did dense knots of gas spiral into the center of the galaxy where they ultimately collected and formed stars?  This question will be examined more closely in Section \ref{starformationefficiency}.

In Figure~\ref{fig:canonical_SFR}, we show the star formation histories for our halos.  These are computed by selecting the particles within $r_{200}$ at $z=0$ and using the stellar age of each particle to compute the implied star formation rate.  This means that the rate shown includes star formation in all progenitor halos (not just the most massive progenitor shown in Figure~\ref{fig:canonical_mass}).  The star formation history 
shows an early burst, as dense, cold gas is rapidly converted into stars, followed by a decline to a steady continuous level of SFR $\sim 3-10 $ \msun\ yr$^{-1}$.  This low-level of star formation reflects both the decreasing gas supply and the decreasing amount of cold gas accretion \citep[e.g.][]{Keres:2005p1111}.  Its value is slightly higher than but roughly consistent with observed levels in the Milky Way of $\sim1$ \msun\ yr$^{-1}$ \citep{Robitaille:2010p1113}.  Interestingly, our lowest mass halo, H54SPM, undergoes a much less pronounced early burst of star-formation, never forming more than 15 \msun yr$^{-1}$, which may be a reflection of a more extended merging period in the first few gigayears of evolution, as evidenced by its accretion history in Figure \ref{fig:canonical_mass}.

\begin{figure*}
\begin{center}
\plotone{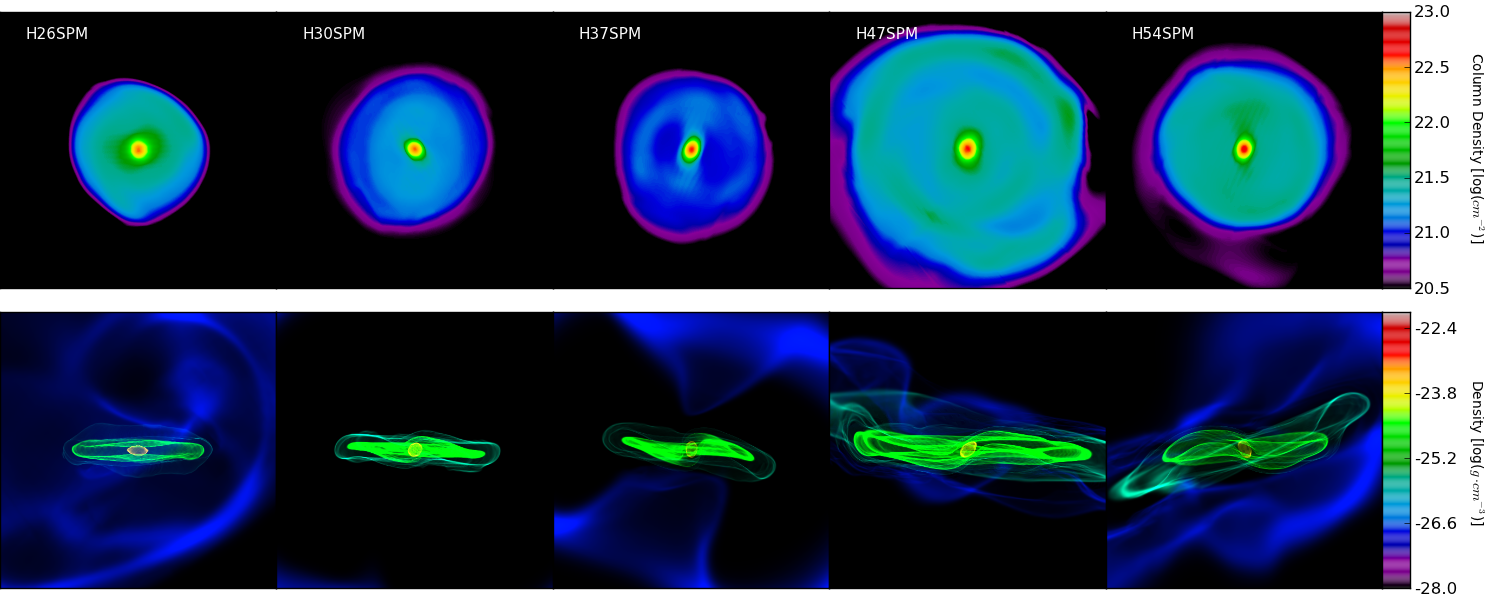}
\plotone{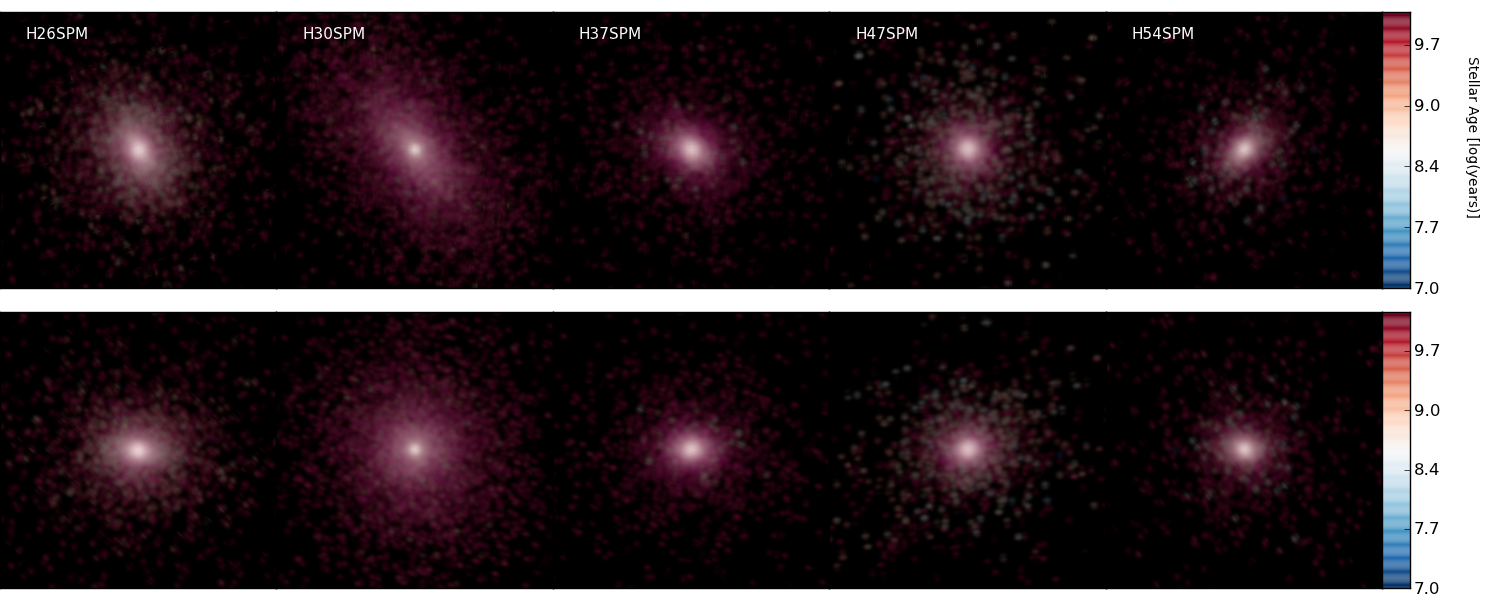}
\end{center}
\caption{We render faceon and edgeon views for each of our five canonical halos at $z = 0$.  The top two rows show the gas component of each galaxy, whereas the bottom two rows display each galaxy's stellar component.  Each postage stamp has a width of 25 kpc.  For the gas, edgeon views show a volume rendering with isocontours of the gas density, whereas faceon views are column density bricks.  For the stars, we use the column density of stars in the faceon and edgeon views respectively.  The color of a star particle is representative of its age where blue represents young stars and red represents old stars according to the color bars on the right.}
\label{fig:canonical_renderings}
\end{figure*}

\subsection{Disk Images}
\label{diskimages}

\subsubsection{Gas}
\label{diskimages_gas}
Each of the five halos produce a gaseous disk at $z=0$.  We use the analysis suite \emph{yt} \citep{Turk:2011p1076} in order to visualize the gaseous and stellar components of these halos, as shown in Figure \ref{fig:canonical_renderings}.   We determine the disk normal vector by computing the net angular momentum of all cold, dense gas (defined as $T < 2 \times 10^4$ K and $\rho > 10^{14}$ \msun Mpc$^{-3}$) within 0.2 $r_{200}$.  We then generate two images for each disk, one side-on and one faceon but each with the same scale of 25 kpc on a side.  These two images are generated in slightly different ways.  For faceon projections, we simply show the gas surface density in units of particles per square centimeter.  For the edgeon image, we carry out a volumetric rendering to show transparent isodensity contours (in detail, we use the gas density to assign a transfer function at each point that consists of a set of narrow Gaussians, each separated by a factor of 10 in density, and each given a different color).  This allows us to show both the disk but also bring out structure in the halo.

These images display gas disks with radii of a few 10's of kpc, slightly smaller than, but approximately typical of present-day late-type systems.  These cold disks are present from early times and are rotationally supported -- we will examine their rotational velocities in more detail in the next section (Section \ref{rotationcurves}).  None of the disks display significant spiral structure at this particular timestep, although all of them seem to possess it at some point in their evolution.  Interestingly, the most massive of the halos, halo 26 does not display as large of a disk as halo 47 despite having nearly two times as much total mass; however Table \ref{tab:halos} reveals that the latter has significantly more cold gas, which is what is plotted here.  The details of satellites and gas accretion tend to dominate the observed behavior of the disk at any point in time, and halo 26 recently accreted some small halos.

\subsubsection{Stars}
\label{diskimages_stars}

In the bottom part of Figure \ref{fig:canonical_renderings}, we render the stellar particle distribution for each halo.  Using the same scale and camera angles as we did for the gaseous components above, we generate stellar surface density plots.  We represent each star particle as a gaussian with a sigma of 200 pc, then step through the volume depthwise, coadding a random sampling of 10\% of all star particles.  The color of a star is determined by its age, where we use a continuous color function of blue through red to represent the log of the age of the star as shown in the colorbar.  This represents stars just 10 Myrs old as pure blue and stars 13 Gyrs old as pure red.  Just as in the gas surface density, the intensity of our stellar renderings is logarithmic with the bulk of the material residing in the inner core of each galaxy.  

All of the stellar halos lack the disk structure we might expect from disk galaxies.  Instead, they are completely dominated by a bulge component.  This happens because star formation occurs most efficiency in the small, dense clumps that merge and lose angular momentum.  Without feedback, the dense gas clumps efficiently deposit gas and stars in the center of the halo, leaving only a gas-poor disk which does not efficiently form stars.  In addition, we do not see significant age gradients for most of the halos.  We note that halo 47 does display the youngest, bluest overall stellar population (particularly relative to halo 30's aging stars), consistent with the star formation histories of Figure \ref{fig:canonical_SFR}.

\subsubsection{Larger gaseous environment}
\label{diskimages_environment}

\begin{figure}
\includegraphics[width=9cm]{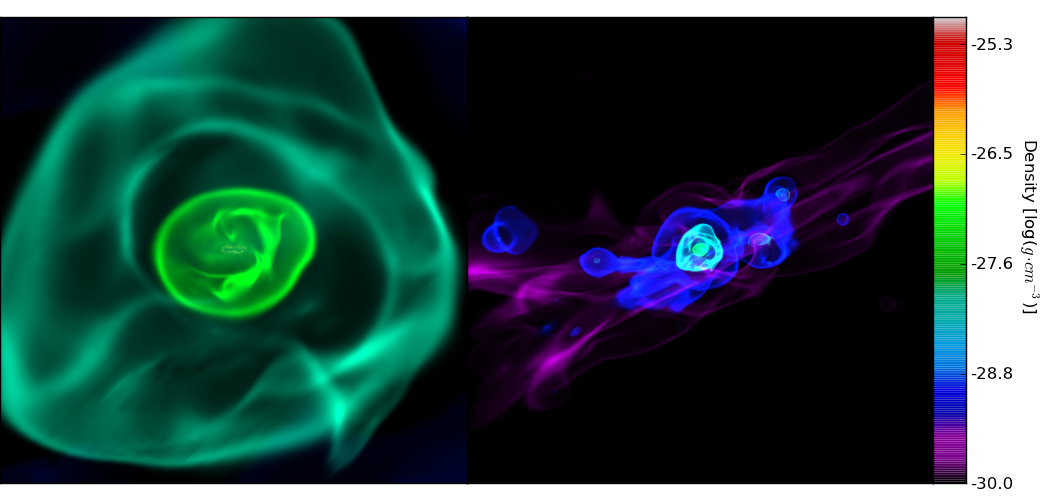}
\caption{We display edgeon volume renderings of Halo 26 at $z = 0$.  Like the edgeon views of the galaxies in Figure \ref{fig:canonical_renderings} these images show structure through the use of isodensity contours; however, unlike Figure \ref{fig:canonical_renderings}, these images show the larger environment around the galaxy with 250 kpc and 2.5 Mpc on a side respectively. }
\label{fig:halo26_large_renderings}
\end{figure}

In addition, we generate large-scale volumetric renderings of halo 26 in order to show its extended halo and immediate environment.  Figure \ref{fig:halo26_large_renderings} is generated in the same way as the edgeon renderings described in Section \ref{diskimages_gas}.  It shows H26SPM over a region with 250 kpc (approximately the virial radius) and 2.5 Mpc on a side respectively.  These large-scale volume renderings (of the side-on disk) show that there is also a gaseous halo, extending out to at least $r_{200} \sim 200$ kpc.  This hot halo gas is approximately spherical at $z=0$, but contains a significant amount of substructure due to ongoing infall and asymmetric accretion.  We do not see gas clumps cooling and condensing out of the smooth, hot gas halo \citep[see also][]{Binney:2009p1106, Joung:2011p1107}.  The larger-scale image shows that the halo is embedded in a set of filaments, along which gas accrete and is typical of the other halos.

\begin{figure*}
\begin{center}
\plotone{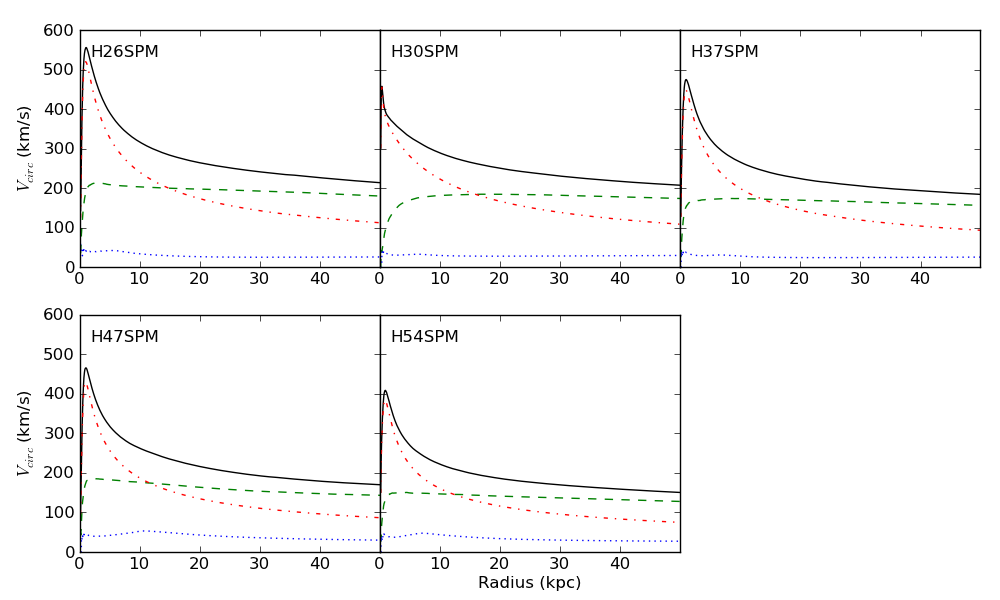}
\end{center}
\caption{We plot the rotation curves for the five simulated halos.  We represent the circular velocity of the gas due to the dark-matter component (green dashed line), the stellar component (red dot-dashed line), the gas component (blue dotted line), and the total of all halo components (black solid line).  The stellar component seems to be driving the cusped rotation curve in the cores of each galaxy, whereas the gas and dark matter profiles appear consistent with observational expectations.  In each system, the total-mass rotation curve is highly cuspy and unlike any observed galaxy, confirming that our simulations reproduce the angular momentum problem.}
\label{fig:canonical_rotation_curves}
\end{figure*}

\subsection{Rotation Curves}
\label{rotationcurves}

In this section, we focus on rotation curves at $z=0$, as these are both directly comparable to observations and also immediately show the mass distribution of the halos.  In Figure \ref{fig:canonical_rotation_curves}, we plot the the rotation curves for each of our five systems.  The curves in each graph represent the equivalent circular velocity for each mass component: dark matter, gas and stars:

\begin{equation}
v_{\rm{circ}}=\sqrt{\frac{G M(\le r)}{r}}
\label{eq:vcirc}
\end{equation}

This figure shows that the rotation curves for each halo are highly peaked in their inner 5 kpc, primarily due to an overabundance of stars in their cores.  This is clearly inconsistent with the nearly flat rotation curves observed in disk systems \citep[e.g.,][]{Courteau:1997p1108}.  The gas contributes negligibly, while the dark matter curve is steeper than expected for an NFW-profile because of contraction driven by the deep potential well of the stellar component, and remains nearly flat to the core, but is secondary to the stellar distribution.

\subsection{Modifications to the Canonical Runs}

In addition to our five canonical runs presented above, we conducted a series of additional simulations in which we systematically varied numerical parameters involving resolution, star formation, supernovae feedback, and gas physics.  We used the initial conditions from halo H26SPM for each of these simulations, so that we could directly compare these results against one of our canonical models.  In the following subsections, we present the results of these various modified runs and for each simulation, examine its star-formation history as well as its rotation curve.  The plots for all of these runs are presented side-by-side in Figures \ref{fig:modified_runs1_SFR}, \ref{fig:modified_runs1_rotation_curves}, \ref{fig:modified_runs2_SFR} \& \ref{fig:modified_runs2_rotation_curves}.  The bulk characteristics of these halos are presented in Table \ref{tab:halos}.  We note that some runs of this halo have a slightly higher virial mass at $z=0$ due to the presence of a fairly large satellite (10\% of the main halo mass), which sits very close to the virial radius -- slight shifts in its position in the various runs can lead to its inclusion in the total mass.

\begin{figure*}[p]
\centering
\includegraphics[width=12cm]{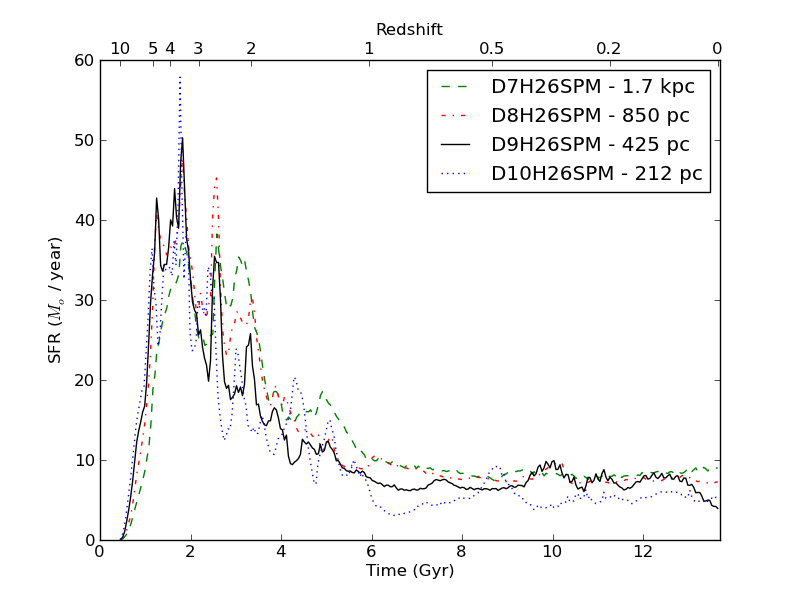}
\caption{We plot the star formation histories for our resolution study.  These runs all use identical initial conditions, those of canonical run H26SPM, which is also presented here as D9H26SPM.  The only difference in each run is the maximum refinement level, that is, the maximum level of spatial resolution achieved which ranges from 212 comoving parsecs to 1.7 comoving kiloparsecs.}
\label{fig:resolution_study_SFR}.
\plotone{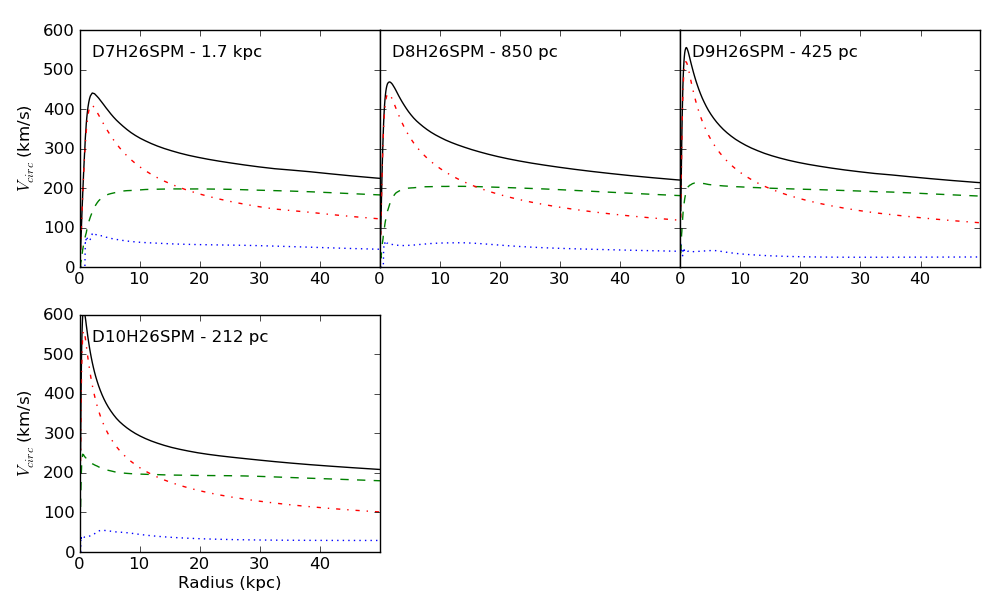}
\caption{We plot the rotation curves for the resolution study.  As in Figure \ref{fig:canonical_rotation_curves}, we represent the circular velocity of the gas due to the dark-matter component (green dashed line), the stellar component (red dot-dashed line), the gas component (blue dotted line), and the total of all halo components (black solid line).  There is a clear trend in these runs that as one increases the resolution of the simulation, one allows a denser cusp of material in the core (primarily in the form of stars).  In turn, the core circular velocity is increasingly driven upward.}
\label{fig:resolution_study_rotation_curves}
\end{figure*}

\subsubsection{Resolution study}
\label{resolution}

Since spatial resolution has been raised as an important issue affecting the angular momentum of the gas \citep[e.g.][]{Mayer:2004p1114}, we conducted three different modified runs, each with a factor of 2 change in the spatial resolution (done by systematically changing the maximum allowed refinement level).  Runs D7H26SP, D8H26SP \& D10H26SP follow the initial conditions of H26SPM (which itself has a resolution of 425 comoving pc) with 1700, 850, and 212 comoving parsecs spatial resolution respectively.  There is no change to the mass resolution in these modified runs.  

Figures~\ref{fig:resolution_study_SFR} and \ref{fig:resolution_study_rotation_curves} show the star formation histories and $z=0$ rotation curves for these runs.  Up to the range we can probe, resolution alone doesn't appear to play a significant role in determining the overall mass distribution in a halo or the conversion of gas into stars.  There is an indication from Figure \ref{fig:resolution_study_SFR} that lower resolution lowers the initial burst (because of the decreased central gas densities) and therefore shifts the peak of star formation to later times; however, the effect is small.  Increasing the spatial resolution actually allows material to condense to an even smaller volume in the cores of galaxies, which increases the inner cusp in the rotation curves of these systems, as seen in Figure \ref{fig:resolution_study_rotation_curves}.  Thus, resolution by itself cannot solve the angular momentum problem.  Perhaps increased resolution when coupled with a more sophisticated star-formation or feedback prescription will produce more realistic galaxies when one begins to resolve star-forming regions on parsec scales.

\subsubsection{Minimum Pressure Support}
\label{pressuresupport}

These simulations cannot resolve star-forming events on parsec scales, so our canonical runs employed an artificial minimum pressure to prevent Jeans fragmentation on smaller scales than we could resolve.  By including this minimum pressure described in \citet{Machacek:2001p1047}, we assured that the Jeans length is always refined by at least 8 cells, and therefore the Truelove criteria was met \citep[see also][]{Truelove:1997p1046, Robertson:2008p1017, Ceverino:2010p1012}.  It has been suggested that artificial disk fragmentation leads to large gas clumps that lose angular momentum via dynamical friction, and hence result in angular momentum loss.  For modified run H26S, we turned off this minimum pressure support.  The results of that run are presented in Figure \ref{fig:modified_runs1_SFR} and \ref{fig:modified_runs1_rotation_curves}.  The results are nearly identical for both measures, indicating that the minimum pressure support did not have a significant effect on the star formation rate or the distribution of gas and stars in our simulated system.

\subsubsection{Star Formation Efficiency}
\label{starformationefficiency}

The star formation model we adopt in this work is based on the Kennicutt-Schmidt relation, but has an efficiency parameter that is not well-constrained.   It has recently been suggested by \citet{Agertz:2010p461} that star formation efficiency is a key parameter controlling the distribution of stars in the disk, and hence the rotation curve.  Our efficiency is already fairly low; however in order to examine this suggestion in more detail, and also to probe the impact of decreasing the efficiency in general, we adopt $\epsilon_{\rm SF} = 10^{-3}$ in run H26SPML.  The star formation history, in Figure~\ref{fig:modified_runs1_SFR}, shows a significant delay to later time, as would be expected, although the net amount of stars produced is nearly identical.  However, this does not translate into a flatter rotation curve, as can be seen in Figure~\ref{fig:modified_runs1_rotation_curves}.  

\begin{figure}
\begin{center}
\plotone{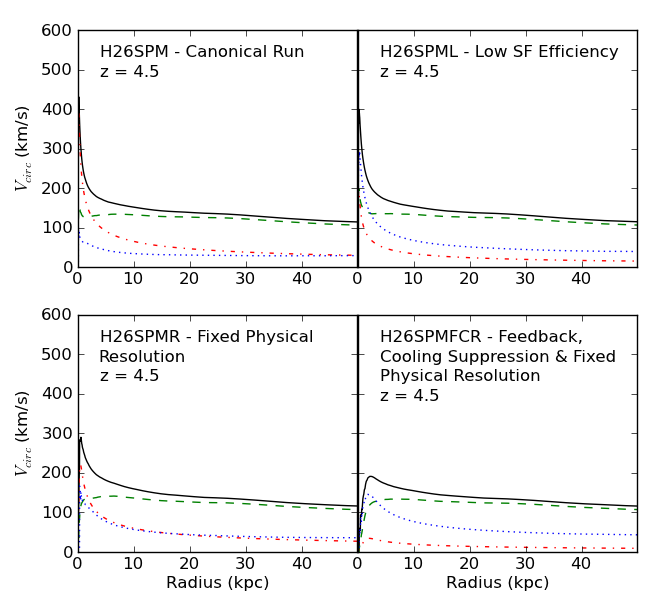}
\end{center}
\caption{We plot the rotation curves for our canonical run (top-left) and the low star-formation run (top-right) at $z = 4.5$.  Both exhibit a cusp of material in the core at this early epoch; however, in the canonical run the cusp is due to stars, whereas the low star-formation run's cusp is almost entirely compressed gas.  This demonstrates that gas is first funneled into the core of a halo even in the absence of significant star formation.  As seen in Figure \ref{fig:modified_runs1_rotation_curves} that in the end, the stellar component ends up dominating the cusp by $z = 0$, regardless of how things look at $z = 4.5$.  In the bottom two panels we also show rotation curves at $z=4.5$ for the fixed physical resolution run and the `everything' run, for comparison.}
\label{fig:lowSFE_rotation_curves}
\end{figure}

To understand this result a little more, we look at the rotation curve for this run, at early times.  Figure~\ref{fig:lowSFE_rotation_curves} shows the rotation curves at $z=4.5$ of this low star formation efficiency run H26SPML compared against our canonical run H26SPM.  During this epoch, much of the halo has formed, but star formation has not yet converted most of the gas into stars.  This plot demonstrates that gas dominates the rotation curve at early times, and yet despite this, the gas clumps have already lost their angular momentum and formed a central cusp of compressed gas.  In runs with higher star formation rates (e.g. the canonical run H26SPM), the gas clumps are partially converted to stars before accreting so the cusp is primarily composed of stars, but the net result is the same -- the clumps lose angular momentum and form a centrally peaked rotation curve. 

Note that it is still possible that star formation efficiency coupled with feedback may be important -- in particular these results are consistent with the idea that low star formation rates combined with sufficient feedback to puff up gas-dominated clumps would suppress angular momentum loss (indeed, the runs in \citet{Agertz:2010p461} generally include feedback and always include a stiff equation of state, $P \sim \rho^2$).  However, we see that by itself, a low star formation efficiency does not change the distribution of matter in the simulated galaxies.

Also shown in Figure~\ref{fig:lowSFE_rotation_curves} are the rotation curves at $z=4.5$ for the fixed physical resolution run (H26SPMR), as well as the simulation including feedback, cooling suppression, and fixed resolution (H26SPMFCR).  As can be seen, the cusp already appears in the fixed resolution run (although not quite as high as in the canonical run), while gas dominates in the H26SPMFCR simulation.  For that run, the additional pressure from feedback/cooling suppression has allowed the infalling clumps to be incorporated in the disk before losing a significant amount of angular momentum.  For the rest of the modified physics runs (not shown), only the cooling suppression simulations differ significantly from the canonical run, and they are similar to the H26SPMFCR simulation, but with a somewhat more pronounced cusp.

\subsubsection{Fixed Physical Resolution}
\label{physicalresolution}

\begin{figure*}
\centering
\includegraphics[width=12cm]{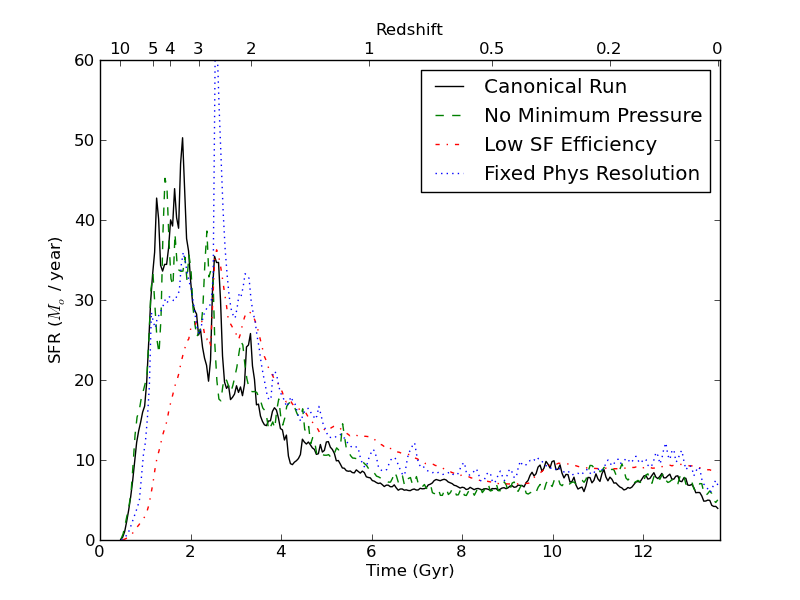}
\caption{We plot the star formation histories for the first three of our modified runs.  These runs all use identical initial conditions, those of canonical run H26SPM, which is also presented here. It appears that these three test runs have little effect on the bulk star formation history of the galaxy, although the simulation with the depressed star formation efficiency delays massive star formation for a gigayear or so.}
\label{fig:modified_runs1_SFR}
\plotone{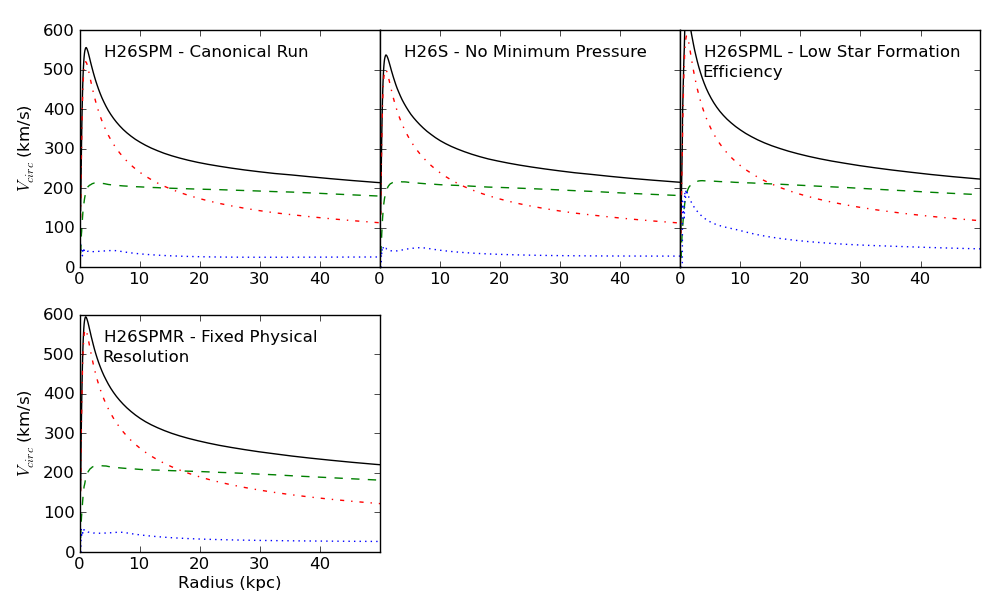}
\caption{We plot the rotation curves for the first three of our modified runs.  Just as in Figures \ref{fig:canonical_rotation_curves} \& \ref{fig:resolution_study_rotation_curves}, we represent the circular velocity of the gas due to the dark-matter component (green dashed line), the stellar component (red dot-dashed line), the gas component (blue dotted line) and the total of all halo components (black solid line).  Like Figure \ref{fig:modified_runs1_SFR}, there is little improvement between these modified runs and their control run H26SPM.  In fact, decreasing the star formation efficiency as in H26SPML actually drives more material into the core by $z=0$.}
\label{fig:modified_runs1_rotation_curves}
\end{figure*}

Our canonical set of simulations used a fixed maximum refinement level, which translates to a best resolution achievable in comoving spatial units.  Another common prescription is to fix the highest resolution as a fixed physical length scale \citep[e.g.][]{Agertz:2010p461}. We carry out a simulation of Halo 26 in which we vary the maximum allowed refinement level so as to keep as close as possible to a physical resolution of 425 pc.  Note that because of the factor-of-two refinement in AMR, this implies discrete resolution changes at various times during the simulation.  The star formation history for this simulation is shown in Figure~\ref{fig:modified_runs1_SFR}.  The lower resolution at early times (for example, at $z=5$, the spatial resolution is four times worse than in the canonical run) results in lower densities and hence a shift of the bulk of the star formation to later times (as in the lowered star formation efficiency run H26SPML).  The rotation curve is shown in Figure~\ref{fig:modified_runs1_rotation_curves}, and again, there is not a significant change in the distribution of mass in the core.

\subsubsection{Supernovae Feedback}
\label{feedback}

As described in the methodology section (Section \ref{method}), we also performed one run with thermal feedback from Type II SN, using a moderate value of $\epsilon_{\rm FB}$, as given in Table~\ref{tab:halos}.  Figure \ref{fig:modified_runs2_SFR} shows the star formation history for this run, and demonstrates that the inclusion of thermal feedback does have some effect.  There is a reduction in the overall burst of star formation at high redshift relative to the canonical run with the same initial conditions.  At late times the star formation rates are similar.  However, in Figure \ref{fig:modified_runs2_rotation_curves} the end result for the rotation curve at $z=0$ is the same as when feedback is not included.  This occurs because the feedback is not strong enough to destroy the infalling clumps and prevent their loss of angular momentum.  It is possible that stronger feedback will change this; however, we delay a systematic examination of feedback prescriptions for a later paper.

\subsubsection{Cooling Suppression Models}
\label{coolingsuppression}

\begin{figure*}
\centering
\plotone{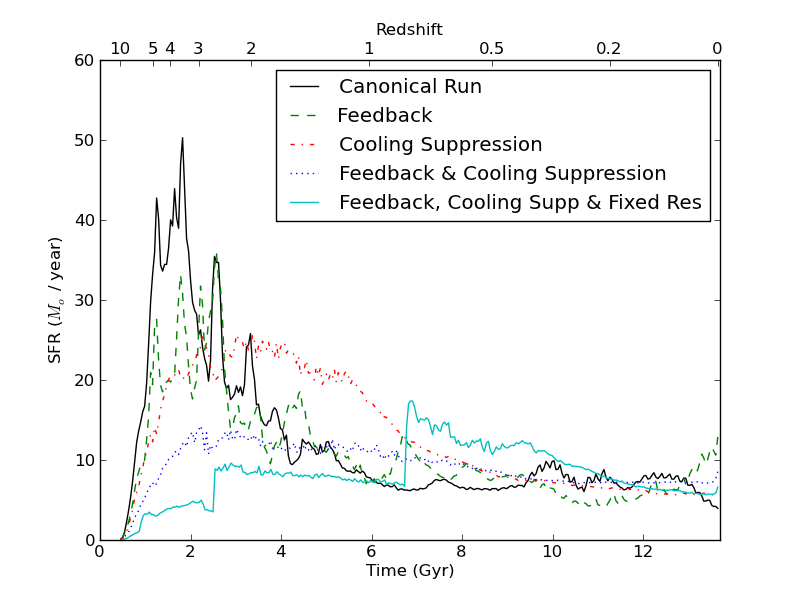}
\caption{We plot the star formation histories for the last four of our modified runs.  These runs all use identical initial conditions, those of canonical run H26SPM, which is also presented here. While feedback and cooling suppression each individually have some effect on lowering overall star formation, together their effects are amplified to keep star formation low throughout the galaxy's lifetime.  The addition of a fixed physical resolution lowers that star formation even further.  Unfortunately, because of the way we have implemented a ``fixed'' physical resolution in \emph{Enzo}, there are discrete jumps in the star formation rate at times when we increase the comoving resolution (e.g. $t\sim2.5$ \& $t\sim7$ gigayears).} 
\label{fig:modified_runs2_SFR}
\plotone{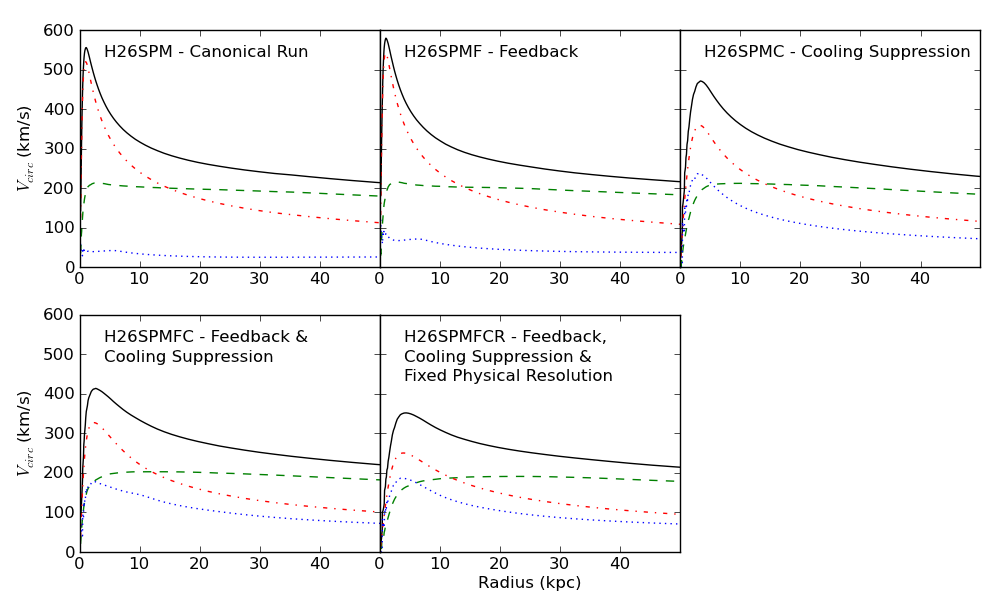}
\caption{We plot the rotation curves for the last four of our modified runs.  Just as in Figures \ref{fig:canonical_rotation_curves} and \ref{fig:resolution_study_rotation_curves}, we represent the circular velocity of the gas due to the dark-matter component (green dashed line), the stellar component (red dot-dashed line), the gas component (blue dotted line) and the total of all halo components (black solid line).  In agreement with Figure \ref{fig:modified_runs2_SFR}, we see that when working in concert the effects of feedback and cooling suppression are intensified to dampen the dense stellar cusp in the interior of our halo.  The addition of the fixed physical resolution further decreases its rotation curve cusp and makes it look much more akin to something we might find from observations.}
\label{fig:modified_runs2_rotation_curves}
\end{figure*}

Many current studies employ a cooling suppression scheme in order to prevent cooling for a short period after a newly formed star is born, allowing the thermal feedback to efficiently operate.  We performed three simulations which integrated cooling suppression: one with cooling suppression alone, one with cooling suppression in addition to thermal feedback, and one with cooling suppression, thermal feedback and fixed physical resolution.  Figure~\ref{fig:modified_runs2_SFR} shows the star formation history of these three simulations.  Interestingly, with just the lone addition of cooling suppression (red dot-dash line), our star formation is highly suppressed at early times, and extends to much later time, also being less bursty compared to the canonical run.  The addition of thermal feedback on top of that further intensifies the effect of cooling suppression and the star formation rate becomes almost constant throughout the simulation at about 10-12 \msun /year.  Finally, the synthesis of feedback, cooling suppression and fixed resolution results in an even lower overall star formation rate.  In this last run, we see a step function in the SFR at two redshifts which correspond to the epochs when we change the allowed maximum refinement level, in order to preserve fixed resolution.  These transitions show up particularly clearly in this case because the cooling suppression operates only in the local cell, and since most of the disk is refined to the maximum level, when this changes, it drastically lowers the efficacy of the cooling suppression, leading to more star formation.  This demonstrates the sensitivity of our star formation results to the chosen parameters of the cooling suppression model.

In Figure~\ref{fig:modified_runs2_rotation_curves}, we show the resulting rotation velocity curves.  We finally have an effect that has a significant impact on the rotation curves.  Even cooling suppression by itself results in a peak rotation rate which is 100 km/s lower than the canonical run.  Adding thermal feedback on top of this decreases the peak even more, to just over 400 km/s, and adding a fixed physical resolution brings it down to about 350 km/s.  This occurs because the cooling suppression model (and the effective feedback that it permits) acts to decrease the density of infalling clumps.  These clumps are then disrupted \emph{before} they can lose a significant amount of angular momentum and so end up rotating at larger radius then they otherwise would have.  We speculate that cooling suppression operates even in the absence of feedback because other forms of heating, such as shock heating and adiabatic compression, can play an important role.  We note that the resulting rotation curves, while much more realistic, are still somewhat too strongly peaked at small radius, a characteristic shared (to a larger or smaller degree) by essentially all simulations that include cooling suppression.

\begin{figure*}
\begin{center}
\plotone{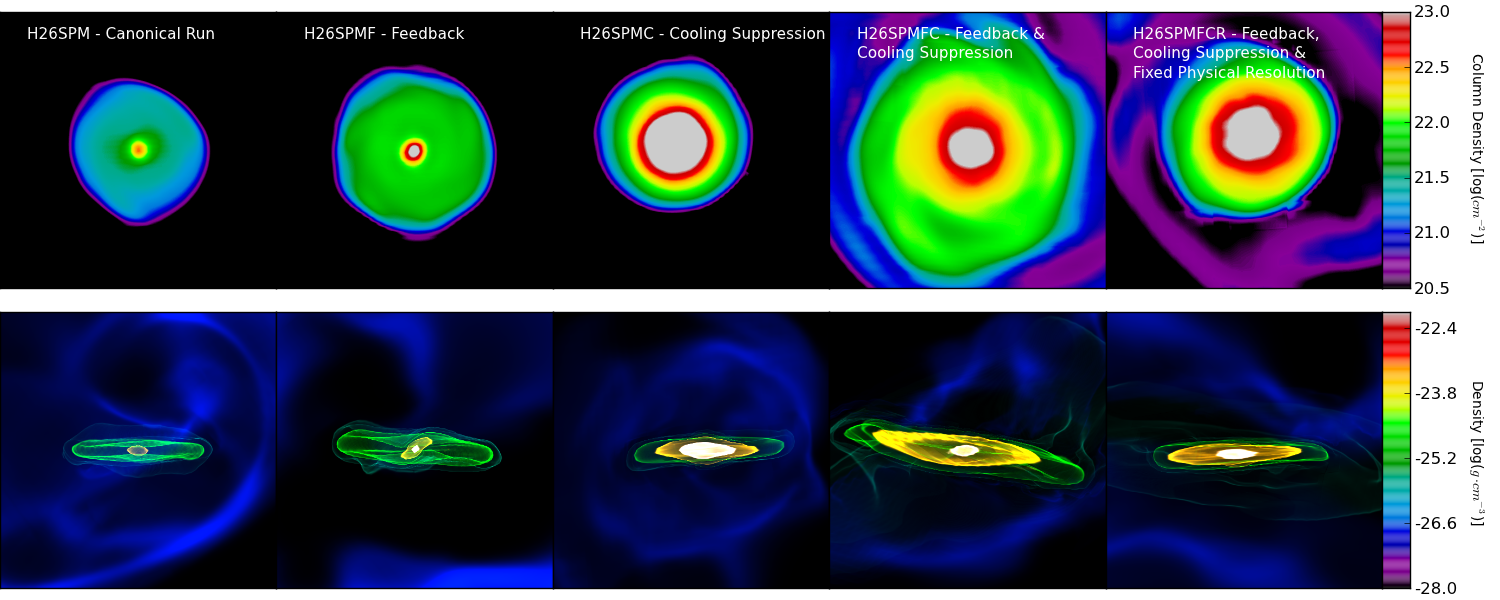}
\plotone{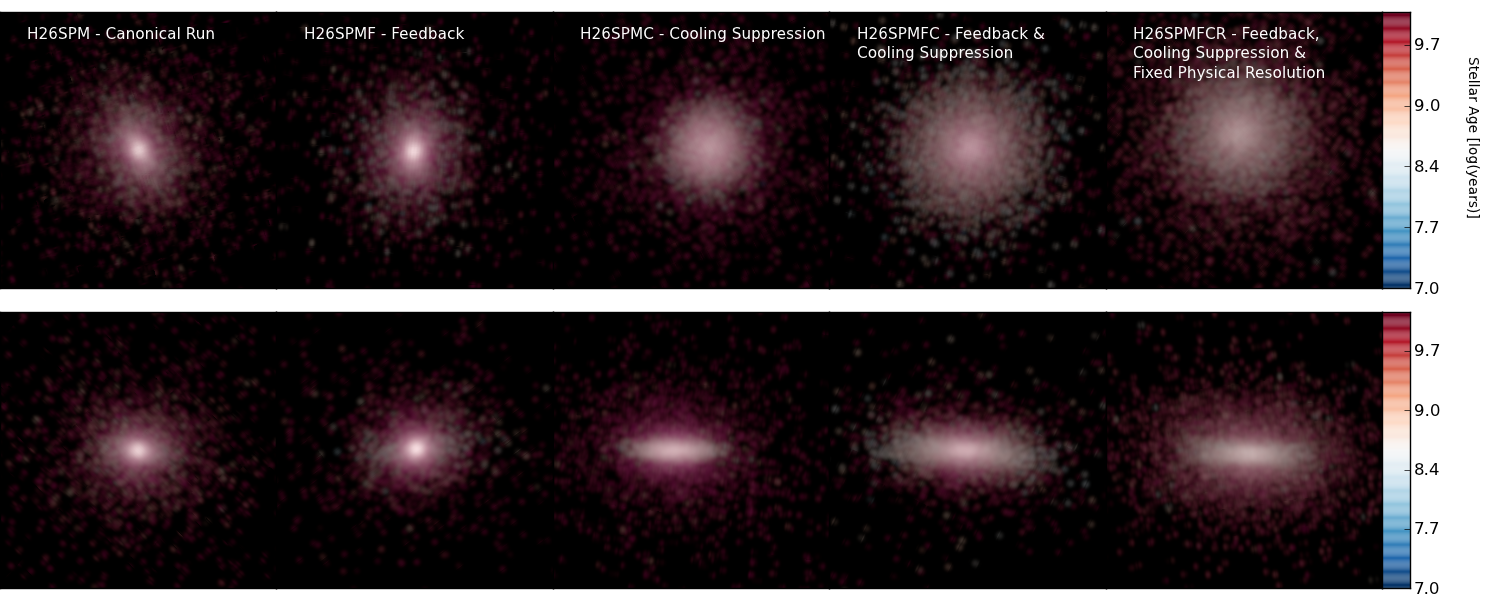}
\end{center}
\caption{We render faceon and edgeon views for the last four of our modified halos at $z = 0$.  The top two rows show each galaxy in gas, whereas the bottom two rows display each galaxy's stellar component.  Each postage stamp has a width of 25 kpc.  For the gas, edgeon views show a volume rendering with isocontours of the gas density, whereas faceon views are column density bricks.  For the stars, we use the column density of stars in the faceon and edgeon views respectively.  The color of a star particle is representative of its age where blue represents young stars and red represents old stars according to the color bars on the right.  Notably, these modified runs produce young stellar disks and have a much higher density of gas in their cores than the canonical runs.}
\label{fig:modified_runs2_renderings}
\end{figure*}

Finally, Figure \ref{fig:modified_runs2_renderings} displays visual renderings of the gaseous and stellar components of the halos, similar to those of the canonical runs in Figure \ref{fig:canonical_renderings}.  The simulations which successfully reduced the buildup of material in the core of the galaxies have larger cold gas densities.  There \emph{is} still a lot of material in the core, but much of it is in compressed cold gas instead of stars.  In the star projections, young stellar disks are present and embedded in halos of older stars.  These disks have no significant bars, but are coaligned with their gas disks, as we might predict for systems of this type.

\section{Discussion}
\label{discussion}

These results demonstrate that it is challenging to generate disk systems with the correct mass distribution.  Without effective feedback, the default outcome is for dense clumps to lose angular momentum and result in centrally-cusped rotation curves.  In particular, the simulation with a very low star formation efficiency nicely demonstrates that this result is fundamentally a dynamical one, and does not depend on whether the clumps are primarily gas, or mostly stars.  As long as they are concentrated, they will lose angular momentum and hence rotational support.  Although this result is not new, and there is a long history of SPH simulations which found this result much earlier, we show it here clearly and systematically using a completely different numerical method (AMR).  Therefore, the result is quite general.

In addition, we went on to systematically vary our numerical parameters and investigate a range of resolution and feedback methods.  We confirmed that only cooling-suppression feedback models are capable of significantly changing the mass distribution and hence the rotation curve.

Cooling suppression models do effectively enhance feedback, although it is unclear how physically meaningful this technique is (see Section \ref{code_coolingsuppression}), and a better approach might be to use high-resolution local models to generate subgrid models \citep[see][for some attempts in this direction]{Yepes:1997p1093, Tasker:2006p1072, Ceverino:2009p1014, Joung:2009p1094}.  

Another constraint is the baryon content of galactic halos: a variety of techniques have been used to infer that the baryon-to-dark-matter ratio in galaxies is much smaller than the cosmic mean \citep[e.g.][]{Moster:2010p1095, Behroozi:2010p1096}, implying that a significant amount of mass has been ejected from galactic systems (or never accreted in the first place).  For example, Milky-Way massed halos only appear to host 20\% of their baryons, with the fraction decreasing rapidly for smaller-mass systems \citep{Behroozi:2010p1096}.  We find that all of our simulations result in very high disk baryon fractions; even the run with feedback and cooling suppression has a baryon fraction of about 63\% of the cosmic mean.  This appears to be a general issue with cosmological galaxy simulations \citep[see also the discussion in][]{AvilaReese:2011p1097}. 

We demonstrate that simulations must be run to $z\sim0$ in order to gauge the efficacy of model parameters at minimizing the effects of the angular momentum problem.  Many past studies \citep[e.g.][]{Ceverino:2009p1014, Joung:2009p1010} have traded off simulation run time for increased resolution in their simulations.  While there are some early indicators of the angular momentum problem like bursts of star formation and peaked rotation curves at redshifts as early as $z=5$, successful results during this epoch do not guarantee successful results at $z\sim0$.  We specifically demonstrate this in the case of the low star formation efficiency run H26SPML, which staved off early bursts of star formation but eventually succumbed to the same fate as runs with a normal star formation efficiency.

\subsection{Comparison to Previous Work}

In agreement with previous work using SPH, we find that unless we include efficient feedback, the resulting systems are dominated by a too-large spheroidal component, and the resulting rotation curve is peaked in the center \citep[e.g.][]{Navarro:1991p1002, Navarro:1997p1098, Weil:1998p1030, Abadi:2003p641, Donghia:2006p1032, Zavala:2008p1099, Piontek:2011p1041}.

A number of reasons have been suggested for this in the past, including purely numerical causes, such as angular momentum transfer between the disk and the hot halo \citep[e.g.][]{Okamoto:2005p1101}, or between cold gas clumps and the hot halo \citep[e.g.][]{Thacker:2000p1040}.  The concern was particularly that SPH simulations might be susceptible to this issue because of smoothing between hot and cold phases.  However, the fact that AMR simulations -- which use a completely different numerical method to solve the fluid equations -- find the same result, is an indication that these effects do not dominate.  

Another suggested source of numerical angular momentum loss is in the form of gravitational instabilities which arise from inadequate resolution of the Jeans length in the disk \citep{Truelove:1997p1046, Robertson:2004p1102}.  We tested this idea by running with and without an additional numerical (``Jeans") pressure designed specifically to ensure that the Jeans length was adequately resolved, finding no difference in our results. 

A third numerical reason is the lack of resolution \citep{Governato:2004p1024, Kaufmann:2007p1103}; however, we specifically test this over the computational range available for us, and find no significant difference from 200-1700 pc.  This is in agreement with the SPH simulations of \citet{Piontek:2011p1041}, who also varied their numerical resolution over a wide range, and found no difference.

There have been a number of recent cosmological AMR simulations which we can compare to.  \citet{Colin:2010p1053} used the \emph{ART} code \citep{Kravtsov:1997p1104} to simulate a halo which is smaller by an order of magnitude (about $10^{11}$ M$_\odot$), finding peaked rotation curves (decreasing as the star formation density criterion was reduced).  Although the halo masses are quite different, the essential result seems to be in agreement.    \citet{Agertz:2010p461} used the \emph{RAMSES} AMR code \citep{Teyssier:2002p1105} to simulate a Milky-Way mass galaxy with similar resolution to that found here.  They argued that a low star formation efficiency by itself was enough to produce nearly flat rotation curves (and that feedback was only efficient if extreme amounts of energy were injected).  We have not been able to confirm the first suggestion -- using a range of low efficiencies for star formation, we find that clumps lose angular momentum at high-redshift, and generate steep rotation curves, whether they are in gas or stellar form.  The efficiency controls when the gas is converted to stars, but has little impact on the distribution of the material.  In this, we are in agreement with previous SPH work \citep[e.g.][]{Weil:1998p1030, Donghia:2006p1032, Piontek:2011p1041}.

Finally, we have found that the only way to significantly decrease the peak of the rotation curve was to introduce a sub-grid model which enhanced the efficiency of stellar feedback.  We briefly explored the cooling suppression model and found this to be effective.  This agrees with a substantial number of SPH simulations which adopt this mechanism \citep{Gerritsen:1997p1039, Thacker:2000p1040, SommerLarsen:2003p1116, Stinson:2006p1023, Governato:2007p1022, Agertz:2010p461, Colin:2010p1053, Piontek:2011p1041, Guedes:2011p1080}.  In addition, \citet{AvilaReese:2011p1097} used AMR simulations (with the \emph{ART} code) and also found cooling suppression to be effective in obtaining approximately flat rotation curves. \citet{Ceverino:2009p1014} also used the \emph{ART} code, but with a different sub-grid model, arguing that a model in which stars are born with significant velocities relative to the nascent gas will feed energy into low-density regions, producing efficient feedback and flat rotation curves, although the simulation is only run to $z \sim 3$.  

\section{Summary}
\label{summary}

In this work, we have carried out adaptive mesh refinement simulations of a sample of Milky-Way sized halos in order to better understand how numerical methodology impacts the content and structure predicted by such models.  We selected a series of five halos ranging in mass from $0.5-1.1 \times 10^{12}$ \msun, each picked to have a quiescent mass accretion history over the last 10 Gyr, with no major mergers, in order to focus on halos that have a high chance of hosting disk-like galaxies.  We simulated the halos with high mass and spatial resolution, including gas, radiative cooling, star formation, and in some runs, feedback.  We also took one halo and resimulated it a number of times, modifying the simulation parameters for each run.  Our primary results are presented below.

\begin{itemize}

\item We find that, without any sort of feedback, all five halos produce rotationally supported gas disks.  They are all, however, dominated by a massive and concentrated stellar spheroid, resulting in rotation curves that peaks at about 500 km/s in the central few kpc.  Therefore, we confirm previous SPH simulation work that also found that dissipation allowed dense clumps to lose angular momentum and produce halos which are too cuspy.

\item For one halo, we vary the comoving spatial resolution from 1700 to 212 pc and find the resulting cuspy halo to be completely robust against resolution change.  Using constant physical resolution, rather than constant comoving resolution also has no effect in reducing the central concentration of mass for halos in the absence of other modified physics.

\item Adding a model of local thermal feedback from Type II SN also has little effect, somewhat suppressing star formation, but with no impact on the $z=0$ rotation curve.  Similarly, varying the star formation efficiency had little effect later than $z\sim4$.   The use of an artificial pressure to ensure that the Jeans length was always resolved also resulted in little change to the star formation history or to the mass distribution.

\item The only modification that did have a significant impact on the overall mass distribution in our halos was to suppress cooling in the vicinity of young stars (more precisely we suppress cooling in the local cell for 50 Myr after a star formation event).  This, combined with the addition of thermal feedback and resolving to a fixed physical resolution, led to a large decrease in the peak of the rotation curve (although still larger than observations indicate in the inner few kpc).

\end{itemize}

To date, the most effective means for staving off the angular momentum problem is to employ cooling suppression in the vicinity of star formation events in order to intensify the effects of feedback.  Most of the recent researchers in this field employ it by default, while at the same time they investigate the effects of other simulation parameters on preventing the angular momentum problem \citep[e.g.][]{Agertz:2010p461, Guedes:2011p1080}.  Our results indicate that the other parameters are secondary to cooling suppression, and they do not seem to work effectively in the absence of it.  We realize that the use of cooling suppression is mostly motivated by its effectiveness in this regard, but we continue to search for other feedback parameterizations which are more physically derived.  

There are several potential alternatives to cooling suppression as a form of feedback in cosmological hydrodynamics solutions of galaxy formation.  Radiative feedback shows some promising results \citep{Kim:2011p1119}, although scaling its effects to a large number of particle sources is a current computational challenge.   Another option is that the cosmic rays produced by supernovae could be used as a means for transporting energy and momentum to the surrounding medium as an additional fluid in a simulation \citep[e.g.,][]{Miniati:2001bz, Jubelgas:2008je}.  Alternatively, modifying the gas in dense regions to have a stiff equation of state \citep{Agertz:2010p461} may help ``puff'' up early collapsing pockets of gas.  Perhaps there is even a redshift-dependent feedback prescription, similar to those used by \citet{SommerLarsen:2003p1116} and \citet{Okamoto:2005p1101}, where it was assumed that the IMF was top-heavy in the distant past resulting in more supernovae and a higher feedback efficiency in that epoch.  We are currently investigating some of these options to be presented in a forthcoming paper.

\acknowledgements

We acknowledge support from NSF grants AST-0547823, AST-0908390, and AST-1008134, as well as computational resources from NASA, the NSF Teragrid, and Columbia University's Hotfoot cluster.  We also thank the \emph{Enzo} and \emph{yt} communities for their helpful discussion and problem-solving on various aspects of the production and analysis of this work.  We benefited from discussions on this topic with Tom Abel, Romeel Dav\'{e}, Erika Hamden, M. Ryan Joung, Sam Leitner, Kristen Menou, Jeff Oishi, Jeremiah Ostriker, Mary Putman, Joop Schaye, David Schiminovich, Sam Skillman, Britton Smith, Stephanie Tonnesen, Matt Turk and John Wise.

\bibliography{paper1.bib}

\begin{thebibliography}{74}
\expandafter\ifx\csname natexlab\endcsname\relax\def\natexlab#1{#1}\fi

\bibitem[{REV(????)}]{REVTEX41Control}
 ????

\bibitem[{08(1)}]{apsrev41Control}
08. 1

\bibitem[{Abadi {et~al.}(2003)Abadi, Navarro, Steinmetz, \&
  Eke}]{Abadi:2003p641}
Abadi, M.~G., Navarro, J.~F., Steinmetz, M., \& Eke, V.~R. 2003, The
  Astrophysical Journal, 591, 499

\bibitem[{Agertz {et~al.}(2010)Agertz, Teyssier, \& Moore}]{Agertz:2010p461}
Agertz, O., Teyssier, R., \& Moore, B. 2010, Monthly Notices of the Royal
  Astronomical Society, 1527, (c) Journal compilation {\copyright} 2010 RAS

\bibitem[{Avila-Reese {et~al.}(2011)Avila-Reese, Col{\'\i}n,
  Gonz{\'a}lez-Samaniego, Valenzuela, Firmani, Vel{\'a}zquez, \&
  Ceverino}]{AvilaReese:2011p1097}
Avila-Reese, V., Col{\'\i}n, P., Gonz{\'a}lez-Samaniego, A., Valenzuela, O.,
  Firmani, C., Vel{\'a}zquez, H., \& Ceverino, D. 2011, eprint arXiv:1103.4422

\bibitem[{Behroozi {et~al.}(2010)Behroozi, Conroy, \&
  Wechsler}]{Behroozi:2010p1096}
Behroozi, P.~S., Conroy, C., \& Wechsler, R.~H. 2010, The Astrophysical
  Journal, 717, 379

\bibitem[{Bigiel {et~al.}(2008)Bigiel, Leroy, Walter, Brinks, de~Blok, Madore,
  \& Thornley}]{Bigiel:2008bs}
Bigiel, F., Leroy, A., Walter, F., Brinks, E., de~Blok, W. J.~G., Madore, B.,
  \& Thornley, M.~D. 2008, The Astronomical Journal, 136, 2846

\bibitem[{Binney {et~al.}(2009)Binney, Nipoti, \&
  Fraternali}]{Binney:2009p1106}
Binney, J., Nipoti, C., \& Fraternali, F. 2009, Monthly Notices of the Royal
  Astronomical Society, 397, 1804

\bibitem[{Brook {et~al.}(2010)Brook, Governato, Roskar, Stinson, Brooks,
  Wadsley, Quinn, Gibson, Snaith, Pilkington, House, \&
  Pontzen}]{2010arXiv1010.1004B}
Brook, C.~B., {et~al.} 2010, ArXiv e-prints

\bibitem[{Bryan \& Norman(1997)}]{Bryan:1997p869}
Bryan, G.~L., \& Norman, M.~L. 1997, eprint arXiv, 10187

\bibitem[{Cen(2011)}]{Cen:2011bp}
Cen, R. 2011, The Astrophysical Journal, 741, 99

\bibitem[{Cen \& Ostriker(1992)}]{Cen:1992p1071}
Cen, R., \& Ostriker, J.~P. 1992, The Astrophysical Journal, 399, L113

\bibitem[{Ceverino {et~al.}(2010)Ceverino, Dekel, \&
  Bournaud}]{Ceverino:2010p1012}
Ceverino, D., Dekel, A., \& Bournaud, F. 2010, Monthly Notices of the Royal
  Astronomical Society, 404, 2151

\bibitem[{Ceverino \& Klypin(2009)}]{Ceverino:2009p1014}
Ceverino, D., \& Klypin, A. 2009, The Astrophysical Journal, 695, 292

\bibitem[{Col{\'\i}n {et~al.}(2010)Col{\'\i}n, Avila-Reese,
  V{\'a}zquez-Semadeni, Valenzuela, \& Ceverino}]{Colin:2010p1053}
Col{\'\i}n, P., Avila-Reese, V., V{\'a}zquez-Semadeni, E., Valenzuela, O., \&
  Ceverino, D. 2010, The Astrophysical Journal, 713, 535

\bibitem[{Courteau(1997)}]{Courteau:1997p1108}
Courteau, S. 1997, The Astronomical Journal, 114, 2402

\bibitem[{D'onghia {et~al.}(2006)D'onghia, Burkert, Murante, \&
  Khochfar}]{Donghia:2006p1032}
D'onghia, E., Burkert, A., Murante, G., \& Khochfar, S. 2006, Monthly Notices
  of the Royal Astronomical Society, 372, 1525

\bibitem[{Eisenstein \& Hut(1998)}]{Eisenstein:1998p1073}
Eisenstein, D.~J., \& Hut, P. 1998, The Astrophysical Journal, 498, 137

\bibitem[{Fall \& Efstathiou(1980)}]{Fall:1980p1033}
Fall, S.~M., \& Efstathiou, G. 1980, Monthly Notices of the Royal Astronomical
  Society, 193, 189

\bibitem[{Gerritsen(1997)}]{Gerritsen:1997p1039}
Gerritsen, J. 1997, Ph.D. thesis, Groningen University, the Netherlands, (1997)

\bibitem[{Gnedin \& Kravtsov(2011)}]{Gnedin:2011ds}
Gnedin, N.~Y., \& Kravtsov, A.~V. 2011, The Astrophysical Journal, 728, 88

\bibitem[{Governato {et~al.}(2007)Governato, Willman, Mayer, Brooks, Stinson,
  Valenzuela, Wadsley, \& Quinn}]{Governato:2007p1022}
Governato, F., Willman, B., Mayer, L., Brooks, A., Stinson, G., Valenzuela, O.,
  Wadsley, J., \& Quinn, T. 2007, Monthly Notices of the Royal Astronomical
  Society, 374, 1479

\bibitem[{Governato {et~al.}(2004)Governato, Mayer, Wadsley, Gardner, Willman,
  Hayashi, Quinn, Stadel, \& Lake}]{Governato:2004p1024}
Governato, F., {et~al.} 2004, The Astrophysical Journal, 607, 688

\bibitem[{Guedes {et~al.}(2011)Guedes, Callegari, Madau, \&
  Mayer}]{Guedes:2011p1080}
Guedes, J., Callegari, S., Madau, P., \& Mayer, L. 2011, eprint arXiv:1103.6030

\bibitem[{Haardt \& Madau(1996)}]{Haardt:1996p1000}
Haardt, F., \& Madau, P. 1996, The Astrophysical Journal, 461, 20

\bibitem[{Joung {et~al.}(2011)Joung, Bryan, \& Putman}]{Joung:2011p1107}
Joung, M., Bryan, G., \& Putman, M. 2011, eprint arXiv:1105.4639

\bibitem[{Joung \& Low(2006)}]{Joung:2006p1038}
Joung, M. K.~R., \& Low, M.-M.~M. 2006, The Astrophysical Journal, 653, 1266

\bibitem[{Joung {et~al.}(2009{\natexlab{a}})Joung, Cen, \&
  Bryan}]{Joung:2009p1010}
Joung, M.~R., Cen, R., \& Bryan, G.~L. 2009{\natexlab{a}}, The Astrophysical
  Journal Letters, 692, L1

\bibitem[{Joung {et~al.}(2009{\natexlab{b}})Joung, Low, \&
  Bryan}]{Joung:2009p1094}
Joung, M.~R., Low, M.-M.~M., \& Bryan, G.~L. 2009{\natexlab{b}}, The
  Astrophysical Journal, 704, 137

\bibitem[{Jubelgas {et~al.}(2008)Jubelgas, Springel, En{\ss}lin, \&
  Pfrommer}]{Jubelgas:2008je}
Jubelgas, M., Springel, V., En{\ss}lin, T., \& Pfrommer, C. 2008, Astronomy and
  Astrophysics, 481, 33

\bibitem[{Katz(1992)}]{Katz:1992p1037}
Katz, N. 1992, The Astrophysical Journal, 391, 502

\bibitem[{Katz {et~al.}(1996)Katz, Weinberg, \& Hernquist}]{Katz:1996p1031}
Katz, N., Weinberg, D.~H., \& Hernquist, L. 1996, The Astrophysical Journal
  Supplement, 105, 19

\bibitem[{Kaufmann {et~al.}(2007)Kaufmann, Mayer, Wadsley, Stadel, \&
  Moore}]{Kaufmann:2007p1103}
Kaufmann, T., Mayer, L., Wadsley, J., Stadel, J., \& Moore, B. 2007, Monthly
  Notices of the Royal Astronomical Society, 375, 53

\bibitem[{Kennicutt(1989)}]{Kennicutt:1989cu}
Kennicutt, R. C.~J. 1989, Astrophysical Journal, 344, 685

\bibitem[{Kere{\v s} {et~al.}(2005)Kere{\v s}, Katz, Weinberg, \&
  Dav{\'e}}]{Keres:2005p1111}
Kere{\v s}, D., Katz, N., Weinberg, D.~H., \& Dav{\'e}, R. 2005, Monthly
  Notices of the Royal Astronomical Society, 363, 2

\bibitem[{Kim {et~al.}(2011)Kim, Wise, Alvarez, \& Abel}]{Kim:2011p1119}
Kim, J., Wise, J., Alvarez, M., \& Abel, T. 2011, eprint arXiv:1106.4007

\bibitem[{Komatsu {et~al.}(2009)Komatsu, Dunkley, Nolta, Bennett, Gold,
  Hinshaw, Jarosik, Larson, Limon, Page, Spergel, Halpern, Hill, Kogut, Meyer,
  Tucker, Weiland, Wollack, \& Wright}]{Komatsu:2009p1070}
Komatsu, E., {et~al.} 2009, The Astrophysical Journal Supplement, 180, 330

\bibitem[{Kravtsov {et~al.}(1997)Kravtsov, Klypin, \&
  Khokhlov}]{Kravtsov:1997p1104}
Kravtsov, A.~V., Klypin, A.~A., \& Khokhlov, A.~M. 1997, The Astrophysical
  Journal Supplement, 111, 73

\bibitem[{Krumholz \& Tan(2007)}]{Krumholz:2007p1115}
Krumholz, M.~R., \& Tan, J.~C. 2007, The Astrophysical Journal, 654, 304

\bibitem[{Machacek {et~al.}(2001)Machacek, Bryan, \& Abel}]{Machacek:2001p1047}
Machacek, M.~E., Bryan, G.~L., \& Abel, T. 2001, The Astrophysical Journal,
  548, 509

\bibitem[{Mayer(2004)}]{Mayer:2004p1114}
Mayer, L. 2004, eprint arXiv:astro-ph/0411476

\bibitem[{Miniati(2001)}]{Miniati:2001bz}
Miniati, F. 2001, Computer Physics Communications, 141, 17

\bibitem[{Mo {et~al.}(1998)Mo, Mao, \& White}]{Mo:1998p1034}
Mo, H.~J., Mao, S., \& White, S. D.~M. 1998, Monthly Notices of the Royal
  Astronomical Society, 295, 319

\bibitem[{Moster {et~al.}(2010)Moster, Somerville, Maulbetsch, van~den Bosch,
  Macci{\`o}, Naab, \& Oser}]{Moster:2010p1095}
Moster, B.~P., Somerville, R.~S., Maulbetsch, C., van~den Bosch, F.~C.,
  Macci{\`o}, A.~V., Naab, T., \& Oser, L. 2010, The Astrophysical Journal,
  710, 903

\bibitem[{Navarro \& Benz(1991)}]{Navarro:1991p1002}
Navarro, J.~F., \& Benz, W. 1991, The Astrophysical Journal, 380, 320

\bibitem[{Navarro \& Steinmetz(1997)}]{Navarro:1997p1098}
Navarro, J.~F., \& Steinmetz, M. 1997, The Astrophysical Journal, 478, 13

\bibitem[{Navarro \& White(1994)}]{Navarro:1994p1004}
Navarro, J.~F., \& White, S. D.~M. 1994, Monthly Notices of the Royal
  Astronomical Society, 267, 401

\bibitem[{Okamoto {et~al.}(2005)Okamoto, Eke, Frenk, \&
  Jenkins}]{Okamoto:2005p1101}
Okamoto, T., Eke, V.~R., Frenk, C.~S., \& Jenkins, A. 2005, Monthly Notices of
  the Royal Astronomical Society, 363, 1299

\bibitem[{Oppenheimer \& Dav{\'e}(2008)}]{Oppenheimer:2008bu}
Oppenheimer, B.~D., \& Dav{\'e}, R. 2008, Monthly Notices of the Royal
  Astronomical Society, 387, 577

\bibitem[{O'Shea {et~al.}(2004)O'Shea, Bryan, Bordner, Norman, Abel, Harkness,
  \& Kritsuk}]{OShea:2004p446}
O'Shea, B.~W., Bryan, G., Bordner, J., Norman, M.~L., Abel, T., Harkness, R.,
  \& Kritsuk, A. 2004, eprint arXiv, 3044

\bibitem[{Piontek \& Steinmetz(2011)}]{Piontek:2011p1041}
Piontek, F., \& Steinmetz, M. 2011, Monthly Notices of the Royal Astronomical
  Society, 410, 2625

\bibitem[{Robertson {et~al.}(2004)Robertson, Yoshida, Springel, \&
  Hernquist}]{Robertson:2004p1102}
Robertson, B., Yoshida, N., Springel, V., \& Hernquist, L. 2004, The
  Astrophysical Journal, 606, 32

\bibitem[{Robertson \& Kravtsov(2008)}]{Robertson:2008p1017}
Robertson, B.~E., \& Kravtsov, A.~V. 2008, The Astrophysical Journal, 680, 1083

\bibitem[{Robitaille \& Whitney(2010)}]{Robitaille:2010p1113}
Robitaille, T.~P., \& Whitney, B.~A. 2010, The Astrophysical Journal Letters,
  710, L11

\bibitem[{Scannapieco {et~al.}(2006)Scannapieco, Tissera, White, \&
  Springel}]{Scannapieco:2006p1118}
Scannapieco, C., Tissera, P.~B., White, S. D.~M., \& Springel, V. 2006, Monthly
  Notices of the Royal Astronomical Society, 371, 1125

\bibitem[{Schaye \& Dalla~Vecchia(2008)}]{Schaye:2008p1045}
Schaye, J., \& Dalla~Vecchia, C. 2008, Monthly Notices of the Royal
  Astronomical Society, 383, 1210

\bibitem[{Sedov(1959)}]{Sedov:1959p1074}
Sedov, L. 1959, Similarity and Dimensional Methods in Mechanics, New York:
  Academic Press, 1959, -1,

\bibitem[{Sommer-Larsen {et~al.}(2003)Sommer-Larsen, G{\"o}tz, \&
  Portinari}]{SommerLarsen:2003p1116}
Sommer-Larsen, J., G{\"o}tz, M., \& Portinari, L. 2003, The Astrophysical
  Journal, 596, 47

\bibitem[{Springel \& Hernquist(2003)}]{Springel:2003p1044}
Springel, V., \& Hernquist, L. 2003, Monthly Notice of the Royal Astronomical
  Society, 339, 312

\bibitem[{Steinmetz \& Muller(1995)}]{Steinmetz:1995p1005}
Steinmetz, M., \& Muller, E. 1995, Monthly Notices of the Royal Astronomical
  Society, 276, 549

\bibitem[{Steinmetz \& Navarro(1999)}]{Steinmetz:1999p635}
Steinmetz, M., \& Navarro, J.~F. 1999, The Astrophysical Journal, 513, 555

\bibitem[{Steinmetz \& Navarro(2002)}]{Steinmetz:2002p1035}
---. 2002, New Astronomy, 7, 155

\bibitem[{Stinson {et~al.}(2006)Stinson, Seth, Katz, Wadsley, Governato, \&
  Quinn}]{Stinson:2006p1023}
Stinson, G., Seth, A., Katz, N., Wadsley, J., Governato, F., \& Quinn, T. 2006,
  Monthly Notices of the Royal Astronomical Society, 373, 1074

\bibitem[{Stone \& Norman(1992)}]{Stone:1992p1117}
Stone, J.~M., \& Norman, M.~L. 1992, The Astrophysical Journal Supplement, 80,
  753

\bibitem[{Tasker \& Bryan(2006)}]{Tasker:2006p1072}
Tasker, E.~J., \& Bryan, G.~L. 2006, The Astrophysical Journal, 641, 878

\bibitem[{Taylor(1950)}]{Taylor:1950p1075}
Taylor, G. 1950, Royal Society of London Proceedings Series A, 201, 159

\bibitem[{Teyssier(2002)}]{Teyssier:2002p1105}
Teyssier, R. 2002, Astronomy and Astrophysics, 385, 337

\bibitem[{Teyssier {et~al.}(2010)Teyssier, Chapon, \&
  Bournaud}]{Teyssier:2010ia}
Teyssier, R., Chapon, D., \& Bournaud, F. 2010, The Astrophysical Journal
  Letters, 720, L149

\bibitem[{Thacker \& Couchman(2000)}]{Thacker:2000p1040}
Thacker, R.~J., \& Couchman, H. M.~P. 2000, The Astrophysical Journal, 545, 728

\bibitem[{Truelove {et~al.}(1997)Truelove, Klein, McKee, Holliman, Howell, \&
  Greenough}]{Truelove:1997p1046}
Truelove, J.~K., Klein, R.~I., McKee, C.~F., Holliman, J.~H., Howell, L.~H., \&
  Greenough, J.~A. 1997, The Astrophysical Journal Letters, 489, L179

\bibitem[{Turk {et~al.}(2011)Turk, Smith, Oishi, Skory, Skillman, Abel, \&
  Norman}]{Turk:2011p1076}
Turk, M.~J., Smith, B.~D., Oishi, J.~S., Skory, S., Skillman, S.~W., Abel, T.,
  \& Norman, M.~L. 2011, The Astrophysical Journal Supplement, 192, 9

\bibitem[{Weil {et~al.}(1998)Weil, Eke, \& Efstathiou}]{Weil:1998p1030}
Weil, M.~L., Eke, V.~R., \& Efstathiou, G. 1998, Monthly Notices of the Royal
  Astronomical Society, 300, 773

\bibitem[{Yepes {et~al.}(1997)Yepes, Kates, Khokhlov, \&
  Klypin}]{Yepes:1997p1093}
Yepes, G., Kates, R., Khokhlov, A., \& Klypin, A. 1997, Monthly Notices of the
  Royal Astronomical Society, 284, 235

\bibitem[{Zavala {et~al.}(2008)Zavala, Okamoto, \& Frenk}]{Zavala:2008p1099}
Zavala, J., Okamoto, T., \& Frenk, C.~S. 2008, Monthly Notices of the Royal
  Astronomical Society, 387, 364

\end{thebibliography}

\end{document}